%% file: conference_3gpp_channel_v5.tex
\documentclass[conference,10pt]{IEEEtran}
\usepackage{url}
\usepackage[utf8]{inputenc}
\usepackage{xcolor}
\usepackage{amsmath}
\usepackage{amssymb}

\usepackage[acronyms,nonumberlist,nopostdot,nomain,nogroupskip]{glossaries}
\usepackage{tablefootnote}
\usepackage{booktabs}

\usepackage{tabularx}

\usepackage{tikz}
\usepackage{pgfplots}
\pgfplotsset{compat=newest} 
\pgfplotsset{plot coordinates/math parser=false} 
\newlength\fheight
\newlength\fwidth
\usetikzlibrary{plotmarks,patterns,decorations.pathreplacing,backgrounds,calc,arrows,arrows.meta,spy,matrix}
\usepgfplotslibrary{patchplots,groupplots}
\usepackage{tikzscale}
\usepackage{hyperref}

\usepackage{cite}

\newif\ifexttikz
\exttikzfalse

\ifexttikz
    \usetikzlibrary{external}
\fi

\usepackage{multirow}
\usepackage{tkz-kiviat}

\usepackage[font=scriptsize]{subcaption}
\usepackage[font=footnotesize]{caption}
\usepackage[capitalise]{cleveref}

\usepackage{physics}

\usepackage{mathtools}

\input{acronyms.tex}
\input{myTikz.tex}

\definecolor{desireRed}{RGB}{230,57,60}%
\definecolor{darkPurple}{RGB}{59,31,43}%
\definecolor{springGreen}{RGB}{37,223,145}%
\definecolor{queenBlue}{RGB}{69,123,157}%
\definecolor{spaceCadet}{RGB}{29,53,87}%

\makeglossaries
\IEEEoverridecommandlockouts

\begin{document}
    
\title{Scalable and Accurate Modeling of\\the Millimeter Wave Channel\vspace{-.2cm}}

\author{\IEEEauthorblockN{Paolo Testolina, Mattia Lecci, Michele Polese, Marco Giordani, Michele Zorzi}
        \IEEEauthorblockA{\small Department of Information Engineering, University of Padova, Italy, email:\texttt{\{name.surname\}@dei.unipd.it}}
        \thanks{This work was partially supported by NIST under Award No. 70NANB18H273. Mattia Lecci and Paolo Testolina's activities were supported by Fondazione CaRiPaRo under the grants ``Dottorati di Ricerca'' 2018 and 2019, respectively.}
}

\makeatletter
\patchcmd{\@maketitle}
  {\addvspace{0.5\baselineskip}\egroup}
  {\addvspace{-1\baselineskip}\egroup}
  {}
  {}
\makeatother

\flushbottom
\setlength{\parskip}{0ex plus0.1ex}

\maketitle

\glsunset{nr}

\begin{abstract}
Communication at millimeter wave (mmWave) frequencies is one of the main novelties introduced in the \gls{5g} of cellular networks.
The opportunities and challenges associated with such high frequencies have stimulated a number of studies that rely on simulation for the evaluation of the proposed solutions.
The accuracy of simulations largely depends on that of the channel model, but popular channel models for mmWaves, such as the \glspl{scm}, have high computational complexity and limit the scalability of the scenarios.
This paper profiles the implementation of a widely-used \gls{scm} model for mmWave frequencies, and proposes a simplified version of the 3GPP \gls{scm} that reduces the computation time by up to 12.5 times while providing essentially the same distributions of several metrics, such as the \gls{sinr} in large scale scenarios.
We also give insights on the use cases in which using a simplified model can still yield valid results.
\end{abstract}


\begin{picture}(0,0)(0,-300)
\put(0,0){
\put(0,30){\small This paper has been accepted for presentation at IEEE ICNC 2020. \textcopyright[2020] IEEE.}
\put(0,20){\small Please cite it as Paolo Testolina, Mattia Lecci, Michele Polese, Marco Giordani, Michele Zorzi, Scalable and Accurate Modeling of}
\put(0,10){\small the Millimeter Wave Channel, IEEE International Conference on Computing, Networking and Communications (ICNC), Big Island, HI, 2020}}
\end{picture}

\section{Introduction}
\label{sec:introduction}
The \gls{5g} of cellular networks is the first to adopt frequencies above 6~GHz to provide access connectivity in mobile scenarios, thanks to the support of carrier frequencies up to 52.6~GHz in 3GPP NR~\cite{38300}, i.e., in the \gls{mmwave} bands.
This portion of the spectrum makes it possible to exploit large chunks of untapped bandwidth, enabling multi-gigabit-per-second data rates to the end-users of 5G networks.
Additionally, at \glspl{mmwave} it is possible to pack a large number of antenna elements in a small form factor, and this allows device manufacturers to embed large antenna arrays also in a smartphone or VR headset~\cite{rangan2017potentials}.
The usage of such high frequencies, however, introduces a set of challenges related to the harsh propagation environment, i.e., the high isotropic pathloss and the sensitivity to blockage~\cite{raghavan2018statistical}.
In particular, the pathloss is proportional to the square of the carrier frequency and is significantly higher at mmWaves than in the sub-6~GHz band, limiting the coverage area of base stations operating in this band.
Additionally, common materials (e.g., bricks and mortar) or the human body block the propagation of millimeter waves, and, consequently, the sudden appearance of an obstacle between the transmitter and the receiver can disrupt the communication or cause wide variations of the available capacity~\cite{rangan2017potentials}.

The peculiarities of the propagation environment at mmWaves have called for the introduction of novel solutions that span multiple layers of the protocol stack, from physical (e.g., with beamforming) to transport and application layers (to make the most out of the massive but erratic capacity available at mmWaves).
In recent years, the research activities on mmWaves have addressed a number of these issues, for example by introducing the support of directional communications~\cite{giordani2018tutorial}, ultra-dense networks~\cite{petrov2017dynamic} and efficient mobility management strategies~\cite{polese2017jsac}.

\begin{figure}
  \centering  
  \includegraphics[width=.85\columnwidth]{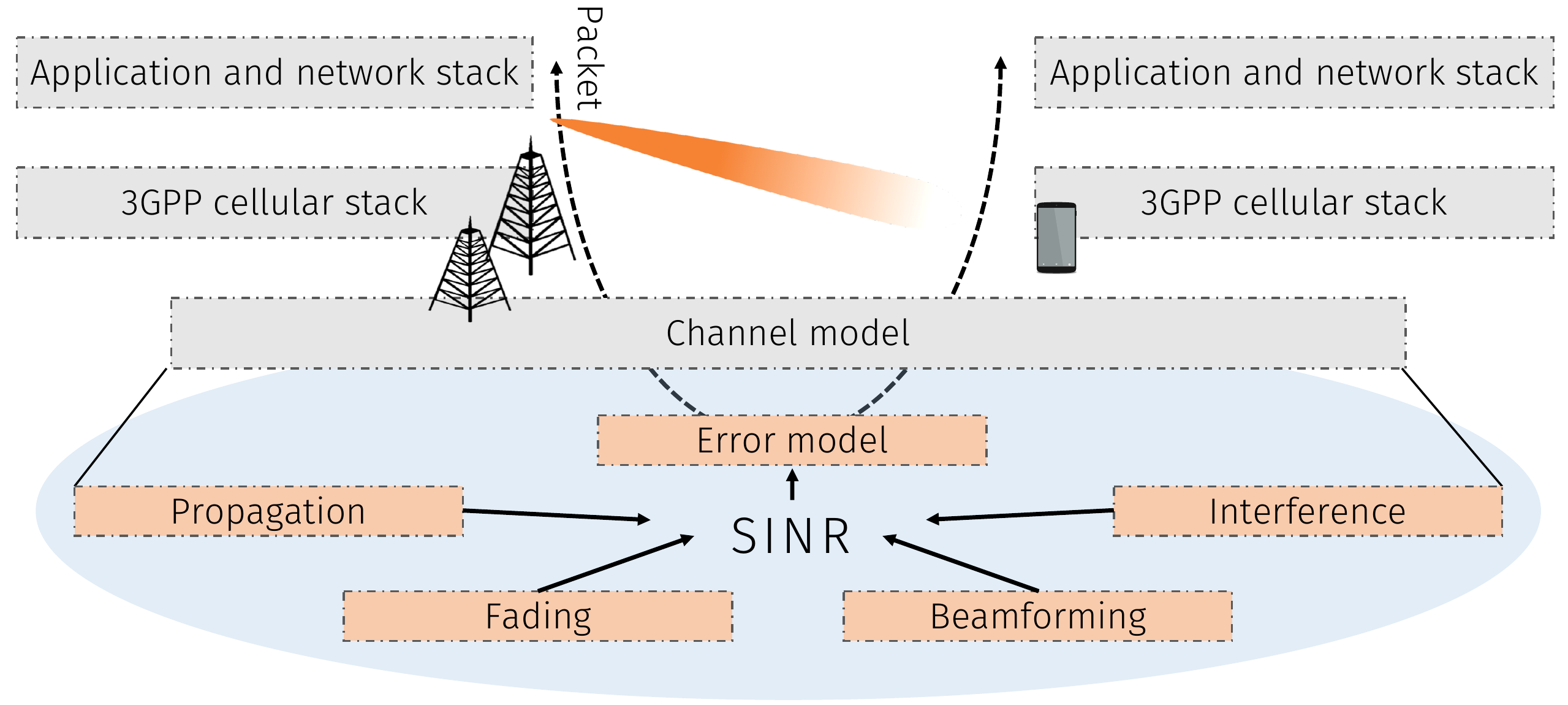}
  \setlength\belowcaptionskip{-.6cm}
  \caption{Typical structure of a system level simulator.}
  \label{fig:sim}
\end{figure}

An important role in this research has been played by simulations.
Indeed, given the early stage of the technological development at mmWaves, there is limited availability of open testbeds and/or real devices and commercial deployments to perform real-world and large-scale experiments.
On the other hand, several open-source or commercial tools have been recently developed to perform link- and system-level simulations~\cite{mezzavilla2018end,sun2017novel,matlab2018toolbox,Patriciello:2019:IML:3321349.3321350}.
Simulators, however, need to accurately model the mmWave channel, to precisely reproduce the effects that the propagation characteristics in these frequency bands introduce throughout the whole protocol stack.
\cref{fig:sim} shows the typical structure of a system-level simulator for mmWave networks: the basic unit is, in general, a packet, whose successful reception at the end device depends on an \gls{sinr} value and an error model that maps it to a packet error rate.
Therefore, a reliable estimation of the \gls{sinr}, which depends on an accurate channel model, is fundamental in the overall performance evaluation.

However, modeling the mmWave channel is one of the most computationally intensive components of a simulation, given that it generally represents large antenna arrays and behaviors in multiple dimensions (i.e., space, time and frequency)~\cite{ferrand2016trends}.
\glspl{scm}, which are popular stochastic models for mmWave frequencies, require the generation of a channel matrix with as many elements as the product of the numbers of transmit and receive antenna elements~\cite{sun2017novel,3gpp.38.901}.
Ray tracers involve the computation of tens to hundreds of multipath components~\cite{jaeckel2014quadriga}.
Some studies are based on simpler channel models, with an abstraction of the beamforming gain and Nakagami or Rayleigh fading~\cite{andrews2017modeling}.
However, as we highlight in our previous work~\cite{polese2018impact}, there exist accuracy and complexity tradeoffs when switching from an \gls{scm} to one of these simpler models.

In this paper, we investigate whether it is possible to simplify the structure of the widely used 3GPP \gls{scm}~\cite{3gpp.38.901} without compromising the accuracy with respect to the original model, used as a baseline.
First, we profile the computational complexity of the model and show that the computations with complex values related to the generation of steering vectors are the main factor that affects the time to generate an instance of the \gls{scm} channel.
We then proceed to simplify these calculations by removing clusters and subpaths (i.e., spatial components of the channel) and comparing the performance of the baseline and simplified models.
We show that some metrics (e.g., the distribution of the \gls{sinr} in a typical 3GPP scenario~\cite{3gpp.38.901}) are not (or only marginally) affected by this simplification, while the channel generation time reduces by up to 12.5 times.
We also highlight the limitations that such a simplification introduces, and give insights on when it may be legitimate to use the simplified version of the model.

The remainder of the paper is organized as follows.
In \cref{sec:channels} we review the main characteristics of mmWave channel models, with a focus on \glspl{scm}.
We then profile the complexity of these models in \cref{sec:profiling}.
\cref{sec:simplification} reports our method and the results of the simplification.
Finally, we provide suggestions for future work in \cref{sec:future_research_directions}.

\section{MmWave Channel Models}
\label{sec:channels}

The accurate modeling of propagation and fading in the mmWave bands has been a topic at the forefront of research in recent years, with multiple measurements and modeling campaigns, such as those described in~\cite{rappaport2013millimeter,gentile2018millimeter}.
The review of mmWave channel models in~\cite{hemadeh2018millimeter} identifies some basic characteristics of the mmWave propagation, i.e., a higher propagation and penetration loss than at sub-6~GHz, the sparsity in the angular domain, the impact of blockage and the clear distinction between \gls{los} and \gls{nlos} states, and the reduced impact of small scale fading.

Channel models for mmWaves can be grouped into three broad classes. 
Quasi-deterministic models~\cite{gentile2018quasi} are extremely accurate in specific scenarios, but need a detailed model of the environment, and are computationally very demanding.
On the other side, models used in analytical studies are based on Nakagami or Rayleigh fading and are usually coupled with a basic sectorized beamforming model~\cite{andrews2017modeling}.
These models simplify the computation of the channel, but are not geometrically related to the scenario, and cannot capture, for example, the spatial dimension of a mmWave channel.

A compromise is usually found using \glspl{scm}, a class of stochastic models that extend the WINNER and WINNER-II models~\cite{winnerII}, can model interactions with beamforming vectors, and have been chosen by the 3GPP for system-level evaluations of 5G networks~\cite{3gpp.38.900}.
In this paper, we focus on the simplification of the 3GPP model for frequencies in the 0.5 to 100~GHz range~\cite{3gpp.38.901}.

An \gls{scm} is given by a propagation loss and a fading model.
The first characterizes the \gls{los} state of the link (probabilistically or with a precise description of the environment) and the average channel gain, with different equations for the \gls{los} and \gls{nlos} conditions~\cite{rappaport2017overview}.

To model fading, the \gls{scm} represents the channel with a matrix $\mathbf{H}$, with $N_{ant,TX}$ rows and $N_{ant,RX}$ columns, that correspond to the transmit/receive antenna elements.
The entry $(i,j)$ of the matrix is given by the combination of $N$ clusters, which model different angular components of the channel between the two transceivers.
The power of each cluster is modeled through an exponential power delay profile, which depends on the delay with which each of the different clusters arrives at the receiver.
Therefore, the \gls{los} path (if present) is the strongest cluster, associated to the minimum delay, followed by several reflections.
Additionally, each cluster can be modeled by the superposition of $M$ subpaths\footnote{The 3GPP specifications refer to subpaths as rays.}, which are distributed with certain statistics around the \gls{aoa} and \gls{aod} of the cluster.

A single realization of the matrix $\mathbf{H}$ depends on the combination of large scale and fast fading parameters.
The first have an impact on the power delay profile, the angular distribution, the relative strength of the \gls{los} component with respect to the \gls{nlos} reflections, and the shadowing.
Large scale parameters generally depend on the scenario that is being modeled.
Fast fading, instead, models small variations in the channel, e.g., the Doppler spread introduced by the user mobility.
The actual parameters may vary in different \glspl{scm}, and are generally expressed through random distributions that fit data collected in measurement campaigns.

\glspl{scm} are popular in the mmWave domain because they have been developed to support beamforming and antenna arrays.
The beamforming gain can indeed be obtained by combining the beamforming vectors for the transmitter and the receiver with the channel matrix $\mathbf{H}$~\cite{akdeniz2014millimeter}.
However, as we will discuss in the next section, the associated computation is one of the most time-consuming parts of the channel generation.

\section{Profiling of the 3GPP mmWave Channel Model}
\label{sec:profiling}

In order to proceed with the simplification of the model, an initial analysis is necessary to understand which are the most computationally demanding steps in the channel generation process.
To remove the dependency on the implementation as much as possible and decouple the model complexity from the implementation inefficiencies, a 3GPP-compliant \cite{3gpp.38.901,3gpp.38.900} network simulator was designed and optimized.
With this tool, we verified experimentally that the channel matrix generation takes up to $90\%$ of the simulation time (the remaining overhead is given by the scenario definition, user mobility, beamforming vector computations, statistical computation, among others).
Although the performance is implementation-dependent, analyzing how the computation of the different parts contributes to the overall simulation time allows drawing general conclusions.

When antenna arrays are considered, the channel model can no longer be expressed through a time-varying scalar impulse response.
Rather, as discussed in \cref{sec:channels}, the channel response is enclosed in a matrix that associates each of the $N_{ant,TX}$ elements of the transmitting array, to each of the $N_{ant,RX}$ elements of the receiving array.
As channel models generate a number of rays coming from different directions, a way to translate such directionality into the definition of the channel matrix is needed.
Considering a narrow-band signal and a small-aperture antenna, the incoming signal seen from the point of view of any given antenna element will be a phase-shifted copy of the original signal.
Steering vectors are used to represent this phase shift over all array elements and are thus composed of complex phase shifts.
Please note that this concept is valid for clusters incoming (or departing) from any direction and for arbitrary arrays.
As \gls{mmwave} use cases are expected to be mainly focused on arrays with tens or hundreds of antennas, the code has been optimized for such scenarios, partially degrading the performance when a small number of antennas (e.g., one at both transmitter and receiver) is used.

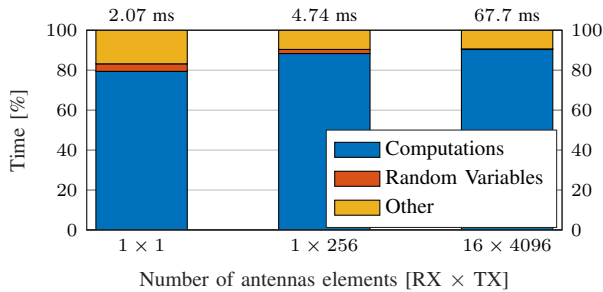
\begin{figure}[t]
  \centering
  \setlength\fwidth{0.75\columnwidth} 
  \setlength\fheight{0.3\columnwidth}
  \input{img/profiling_bars}
  \setlength\belowcaptionskip{-.6cm}
  \caption{Results from the profiling of the 3GPP channel model described in~\cite{3gpp.38.901}, for different square antenna arrays at the transmitter and the receiver.
  We report the percentage of different tasks related to the channel matrix generation, and the absolute execution time above each bar.
  The \emph{Computations} term only includes operations related to steering vectors, whose  complexity increases with the channel matrix's size.
  As the size of the channel matrix increases, operations with complexity proportional to its size dominate over the overall generation time.}
  \label{fig:profiling_bars}
\end{figure}%

Our profiling highlights that the computations related to steering vectors and their combination are the most time-consuming part of the generation of an instance of the matrix representing the 3GPP channel, as reported in the \emph{Computations} entry in \cref{fig:profiling_bars}.
For a channel with a single element (i.e., $\mathbf{H} \in \mathcal{C}^{1 \times 1}$), the computation takes 79.42\% of the time.
This percentage increases up to 90.38\% for the largest antenna array configuration we consider (i.e., $16 \times 4096$).\footnote{Notice that the dependence between the number of antenna elements and the \emph{Computations} entries is not linear, as MATLAB introduces optimizations for large matrices.}
The generation of random variables, such as the cluster powers, the delays, the sub-paths' angles, the phase shifts, and the \gls{aoa}/\gls{aod} coupling, is instead negligible, particularly when large arrays are considered (0.27\% for the $16 \times 4096$ configuration).
On the contrary, the code overhead, composed of sub-routine calls and all other operations, is significant and does not considerably depend on the array size.

As the \emph{Computations} entry is related to the generation and combination of the steering vectors, it is proportional to the number of clusters and subpaths that are generated and combined: the richer the channel, the slower this computation.
For example, as reported in \cref{fig:profiling_bars}, a channel with a single entry would take up to $1.64$~ms to perform the computations for a total of $2.07$~ms, while for the largest antenna configuration, yielding the largest channel matrix, computations take $61.19$~ms for a total of $67.7$~ms. 

\begin{figure*}
  
\begin{subfigure}[t]{.48\textwidth}
  \centering
  \setlength\fwidth{0.8\columnwidth} 
  \setlength\fheight{0.3\columnwidth}
  \input{img/chanGenTime_poster}
  \caption{Computation time}
  \label{fig:chanGenTime_poster}
\end{subfigure}
\hfill
\begin{subfigure}[t]{.48\textwidth}
  \centering
  \setlength\fwidth{0.8\columnwidth} 
  \setlength\fheight{0.3\columnwidth}
  \input{img/speedup}
  \caption{Performance gain}
  \label{fig:speedup}
\end{subfigure}
  \setlength\belowcaptionskip{-.3cm}
  \caption{Computation time required to generate an instance of the channel matrix (\subref{fig:chanGenTime_poster}) and performance gain introduced by the simplification (\subref{fig:speedup}), as a function of the number of antenna elements at the \gls{gnb}, for different configurations at the \gls{ue} and different combinations of simplification parameters for the channel.}
  \label{fig:computation_performance}
\end{figure*}
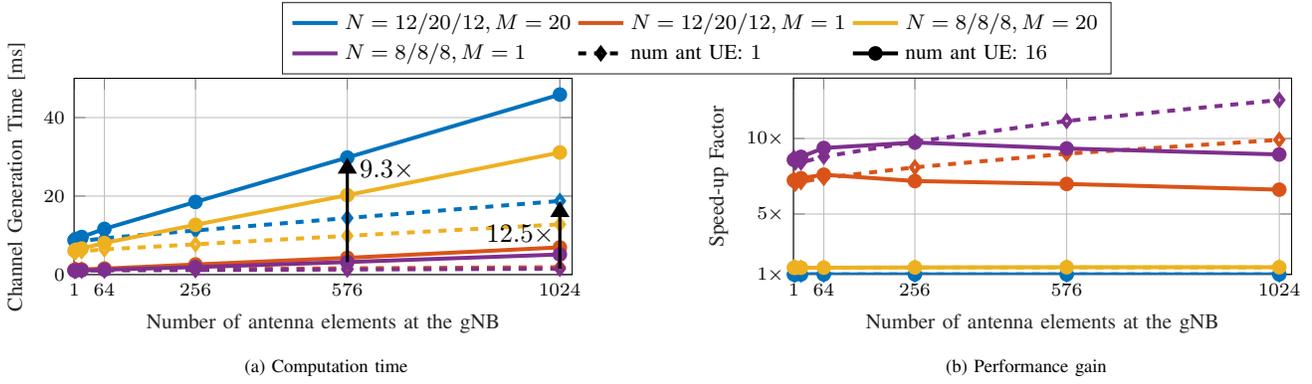

\section{Channel Simplification}
\label{sec:simplification}
As shown by the analysis in \cref{sec:channels}, reducing the number of clusters and subpaths can be beneficial in terms of simulation time.
Nevertheless, changes in the channel model may affect or even compromise its reliability, depending on the application.
In this section, we analyze the effects of the simplification on the network statistics obtained from the network simulator described in \cref{sec:profiling}.
The models with the original number of clusters and sub-paths were considered as baselines with which to compare the effects of the simulations, i.e., we do not provide in this paper a direct comparison with ground truth measurements.
Thanks to the flexibility of our simulator, it was possible to perform the tests on different configurations of the 3GPP channel model while keeping the same settings for the cellular scenario.
For this study, a 3GPP-compliant \gls{uma} downlink scenario is considered~\cite{3gpp.38.901}.
Similar results can also be obtained for other scenarios.

The 3GPP channel model differentiates among three states of the channel, namely \gls{los}, \gls{nlos}, and \gls{o2i}.
Different channel states correspond to a different number $N$ of clusters, whereas the number of sub-paths per cluster $M=20$ is kept fixed for all propagation conditions.
As per~\cite{3gpp.38.901}, $N_{LoS}=12$ clusters are present in \gls{los}, $N_{NLoS}=20$ in \gls{nlos}, and $N_{O2I}=12$ in \gls{o2i} channel conditions, in short $N=12/20/12$.

We followed two complementary simplification strategies: on one hand, reducing the number of clusters, and on the other hand, reducing the number of sub-paths per cluster. 
Indeed, it was possible to vary the latter from $M=20$ to $M=1$, corresponding to a cluster with only the main path and no sub-paths.
On the contrary, in~\cite{3gpp.38.901}, the azimuth and elevation angle spreads are specified only for some specific cluster configurations, and cannot be trivially interpolated to extract the parameters for the configurations that are excluded from the model. Therefore, this limits the extension of our simplification to clusters.
Specifically, the channel model was complete enough to allow only a maximum reduction down to $N_{LoS}=8$, $N_{NLoS}=8$, $N_{O2I}=8$, besides the default one.

Considering the different contributions to the computational complexity discussed in \cref{sec:profiling}, the speed-up factor should be proportional to the reduction of the overall number of clusters and/or sub-paths.
However, several additional aspects need to be taken into account, making the dependence on the number of clusters and sub-paths not necessarily linear.
Particularly, decreasing $N$ for one channel state will contribute proportionally to the number of users that are in that propagation condition.
In the considered scenario, which follows the specifications in~\cite{3gpp.38.901}, $80\%$ of the users, being indoor, are in \gls{o2i} conditions, making $N_{O2I}$ the most significant term to reduce.
Moreover, depending on the implementation and on the initial access policy, one may need to consider only the users who are connected to a \gls{gnb}, adding a further layer of complexity to these considerations.
In our simulator, the attachment is purely based on the combination of pathloss and shadow fading, and the channel is computed only for the users that successfully connect to a \gls{gnb}.

We evaluated the various $(N,M)$ configurations for different array sizes, two for the \glspl{ue} and five for the \glspl{gnb}, to test our approach in multiple settings.
Array sizes were chosen following typical values found in the literature, and scenarios with both single- and multi-antenna \glspl{ue} were tested.
In \cref{fig:chanGenTime_poster,fig:speedup}, the generation time and speed-up factors with respect to the baseline configuration $N=12/20/12$ are shown.
The generation time of a single channel matrix was reduced by a factor up to $12\times$ for a single-antenna \gls{ue}, going from $18.75$~ms to $1.49$~ms.
Note that, according to the aforementioned considerations on the channel state distribution, the speed-up factor is not necessarily proportional to the reduction of the number of clusters.

We evaluated the effects of the simplifications on (i) the narrowband \gls{sinr}, given its relation with channel capacity; (ii) the wideband \gls{sir}; and (iii) the distribution of the singular values of the channel matrix, to show how spatial multiplexing is affected by our channel simplification.
Defining $P_{rx}$, $P_{N}$ and $I_{tot}$ as the powers of the received signal, the noise and the interfering signals, respectively, the narrow-band SINR, computed after the optimal SVD-based SISO beamforming, can be expressed~as
\begin{equation}
  \Gamma = \frac{P_{rx}}{P_{N} + I_{tot}}
\end{equation}
The wide-band SIR is defined as
\begin{equation}
  \xi(f) = \frac{\qty| H_{rx}(f) |^2} {\qty| \sum_{i=1}^{N_{interf}} H_{interf,i}(f) |^2},
\end{equation}
where $H_{rx}(f)$ is the receiver's channel frequency response and $H_{interf,i}(f)$ are the channel frequency responses from the $N_{interf}$ interfering base stations to the receiver.

For the wideband case, following~\cite{itu-1407}, we consider two metrics that 
measure the impact of fading on the performance of the system. The \gls{lcf} is defined as the fraction of \gls{ofdm} subcarriers\footnote{In our scenario, we consider a total bandwidth of 100 MHz, with subcarrier spacing equal to 60 kHz, as specified by the 3GPP for calibration at 30 GHz~\cite{3gpp.38.901}.} in which the \gls{sir} 
$\xi(f)$ (as a function of frequency) crosses a given threshold $\xi_{th}$ 
in the upward (or equivalently downward) direction. The \gls{afbw} is defined as the average width (in kHz) of contiguous chunks of 
the overall bandwidth for which the envelope of $\xi(f)$ stays below a 
given threshold $\xi_{th}$.

\begin{figure}[t]
  \centering
  \setlength\belowcaptionskip{-.3cm}
  \setlength\fwidth{0.75\columnwidth} 
  \setlength\fheight{0.18\columnwidth}
  \input{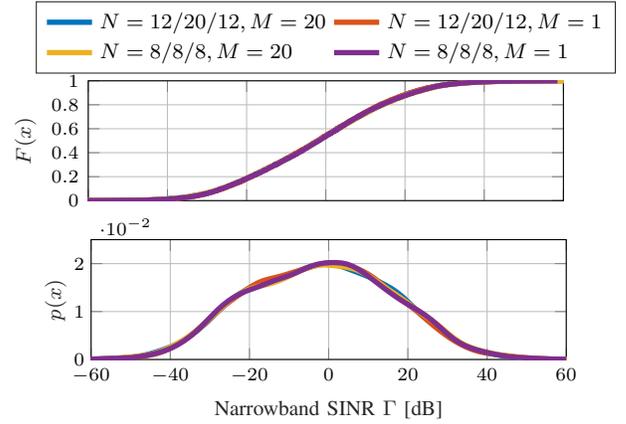}\\
  \input{img/3gpp_8x8_4x4_pdf}
  \caption{Cumulative Distribution and Probability Density Functions of the narrowband SINR $\Gamma$ of a scenario composed of \glspl{ue} and \glspl{gnb} with $16$ and $64$ antenna elements, respectively.}
  \label{fig:3gpp_8x8_4x4}
\end{figure}
\begin{figure}[t]
  \centering
  \setlength\belowcaptionskip{-.3cm}
  \setlength\fwidth{0.75\columnwidth} 
  \setlength\fheight{0.27\columnwidth}
  \input{img/lcr_afd_manyAnt_svd_AFD}
  \caption{\glsreset{afbw}\Gls{afbw} vs $\xi_{th}$ for a scenario with \glspl{ue} equipped with $16$ antenna elements, and \glspl{gnb} with 64 antenna elements.}
  \label{fig:afd_manyAnt_svd}
\end{figure}
\begin{figure}[t]
  \centering
  \setlength\belowcaptionskip{-.3cm}
  \setlength\fwidth{0.75\columnwidth} 
  \setlength\fheight{0.27\columnwidth}
  \input{img/lcr_afd_manyAnt_svd_LCR}
  \caption{\glsreset{lcf}\Gls{lcf} vs $\xi_{th}$ for a scenario with \glspl{ue} equipped with $16$ antenna elements, and \glspl{gnb} with 64 antenna elements.}
  \label{fig:lcr_manyAnt_svd}
\end{figure}
\begin{figure}[t]
  \centering
  \setlength\belowcaptionskip{-.3cm}
  \setlength\fwidth{0.75\columnwidth} 
  \setlength\fheight{0.27\columnwidth}
  \input{img/eigenValues}
  \caption{Mean Singular Value Ratio of channel matrices for a scenario with \glspl{ue} equipped with $16$ antenna elements, and \glspl{gnb} with $64$ antenna elements.
  Only the baseline and the most extreme simplification are shown.}
  \label{fig:eigenValues}
\end{figure}
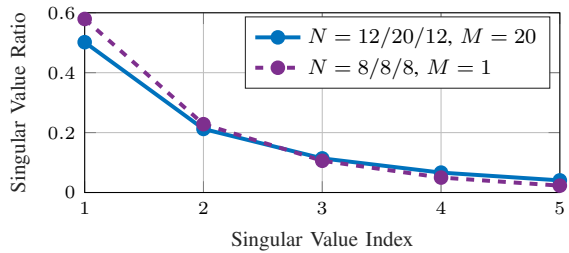

Results show that narrow-band statistics are not affected by the channel simplification (\cref{fig:3gpp_8x8_4x4}), whereas the wide-band ones are only slightly affected by it.
It is interesting to notice how, considering $N=12/20/12,\, M=20$ as the baseline, removing clusters almost does not affect the \gls{afbw} (\cref{fig:afd_manyAnt_svd}) while, on the contrary, the removal of sub-paths does not significantly affect the \gls{lcf} (\cref{fig:lcr_manyAnt_svd}).
From \cref{fig:eigenValues} it can be noted that the mean ratio of the singular values of the channel matrices, while being the most diverging metric shown, still does not significantly differ from the baseline. The first singular value of the baseline model, however, is 14\% smaller than that of the simplified channel. 
In any case, differences between the baseline and the most simplified channel (i.e., $N=8/8/8, M=1$) only have a minor effect on most metrics while speeding up the simulation by a factor of $10$.

Thus, as shown by these results, reducing the number of clusters and sub-paths to the minimum allowed by the parameters found in~\cite{3gpp.38.901} does not significantly change the system performance, while obtaining significant reduction of the computations.
Unfortunately, it is not possible to push the simplification even further, while following the constraints of the parameters in the 3GPP specifications.



\section{Future Research Directions} 
\label{sec:future_research_directions}

As discussed in~\cref{sec:introduction}, modeling the channel at \gls{mmwave} frequencies is a task of primary importance for the proper design and validation of communication protocols in wireless networks. However, the profiling of the 3GPP model for \glspl{mmwave} reported in this paper highlights how complex channels may require a very significant computational overhead. This paper has presented a possible simplification for the 3GPP channel model, which, with respect to the full-fledged 3GPP model, manages to reduce the computational complexity while being accurate for a number of metrics in 3GPP scenarios. Nonetheless, the simplification was constrained by the structure of the 3GPP model itself, as discussed in Sec.~\ref{sec:simplification}.

Future research should therefore focus on how it is possible to further improve the computational complexity of mmWave channels, to quickly generate \gls{mmwave} channel instances even with limited computational resources, without compromising the accuracy.
It is thus necessary to close the gap between models that are too coarse, and therefore not necessarily reliable, and those that are too detailed, and therefore impractical at the system level. 
In particular, the following aspects deserve further investigation.
\smallskip

\smallskip
\emph{a) Impact of channel simplifications on higher layers:}
In \cref{sec:simplification}, we evaluated the effects of channel simplifications on link-level metrics, including narrowband \gls{sinr}, wideband SIR, \gls{lcf}, and  \gls{afbw}. We will extend our future analyses to end-to-end performance metrics, including per-user throughput, latency, or application packet reception ratio. Indeed, as discussed in~\cite{polese2018impact}, different fading models, as well as channel implementations, have different implications on higher-layer metrics, e.g., network latency, when considering the complex interactions with the full protocol stack (e.g., with the congestion control provided by TCP).

Additional research efforts should therefore be devoted to identifying which level of detail is most adequate (i.e., provides accurate results while minimizing the complexity) considering the performance of end-to-end systems.
In this context, system-level simulators, e.g., ns-3~\cite{henderson2008network}, feature a complete TCP/IP stack and thus represent a promising tool to perform accurate simulations.
\smallskip

\smallskip
\emph{b) Impact of  channel simplifications on directionality}:~
In the \gls{mmwave} context, multi-antenna systems have emerged as a solution to combat the severe path loss experienced at high frequencies.
As we demonstrated in \cref{sec:simplification}, minimizing the number of spatial components of the channel has minimal implication on a number of metrics, such as the \gls{sinr}, and significantly reduces the channel computation time. However, it also reduces the multipath components of the channel, thereby preventing the model from being used for the evaluation of techniques that, in reality, exploit channel sparsity to form multiple simultaneous beams in independent angular directions (e.g., hybrid beamforming).

Channel abstractions should therefore also address the trade-off that arises between making the channel more computationally efficient and accurate while preserving the multipath characteristics of \gls{mmwave}~signals.

\smallskip
\emph{c) Simplification of quasi-deterministic channels:}
Besides \glspl{scm} (e.g., the 3GPP channel model in~\cite{3gpp.38.901}), more accurate channel measurements were formalized in quasi-deterministic ray-tracing models~\cite{jaeckel2014quadriga,gentile2018quasi}. These models are much more complex than \glspl{scm} due to the massive number of operations that have to be performed to compute each instance of the channel.
In this regard, more substantial simplifications may need to be introduced in order to make channel models tractable at the system level.

\smallskip
\emph{d) Towards more accurate channel modeling:}
Although existing channel models provide some insights into the propagation characteristics of \glspl{mmwave} in cellular environments, more research is needed to capture the nuances of the propagation and fading in the \gls{mmwave} scenario.
 In particular, future measurement campaigns should capture the following aspects.
 \begin{itemize}
   \item \emph{Second order statistics}. Due to the lack of temporally and spatially correlated channel measurements in the \gls{mmwave} band, few and limited studies analyze the channel autocorrelation function. As a consequence, it is currently not possible to develop accurate statistical models for mobility-related and/or multi-connectivity scenarios, thereby preventing researchers from making a clear assessment of how the channel dynamics impact the overall network performance.

   \item \emph{Doppler spread (with directional antennas)}. It has been shown that directional transmissions affect the power angular profile which in turn affects the Doppler spread. Such an effect has not yet been characterized by currently available channel measurements.

   \item \emph{Propagation characteristics}. Ground reflection plays an important role at both small-scale and large-scale levels, especially in vehicular deployments, but is generally neglected. Human body blockage and the effect of mobile terminal rotation have been modeled in detail in many papers, although their effects are often overlooked. Finally, an indication of the absolute number and the corresponding distributions of the multipath components of the channel has yet to encounter general agreement.

   \item \emph{Modeling of dynamic scenarios.} The channel parameters are typically derived from measurements in indoor and cellular settings, while measurements in more dynamic scenarios, e.g., in a vehicular context or considering aerial user terminals like drones,   are still lacking.
   
 \end{itemize}

Future research activities should extend the available channel models to account for the aforementioned missing elements, without compromising the computational complexity.

\bibliographystyle{IEEEtran}
\bibliography{bibl.bib}

\end{document}

%% file: acronyms.tex
\newacronym{3gpp}{3GPP}{3rd Generation Partnership Project}
\newacronym{adc}{ADC}{Analog to Digital Converter}
\newacronym{afbw}{AFBW}{Average Fading Bandwidth}
\newacronym{5g}{5G}{5th generation}
\newacronym{aimd}{AIMD}{Additive Increase Multiplicative Decrease}
\newacronym{am}{AM}{Acknowledged Mode}
\newacronym{amc}{AMC}{Adaptive Modulation and Coding}
\newacronym{aqm}{AQM}{Active Queue Management}
\newacronym{awgn}{AGWN}{Additive White Gaussian Noise}
\newacronym{balia}{BALIA}{Balanced Link Adaptation}
\newacronym{bdp}{BDP}{Bandwidth-Delay Product}
\newacronym{bf}{BF}{Beamforming}
\newacronym{cc}{CC}{Congestion Control}
\newacronym{cdf}{CDF}{Cumulative Distribution Function}
\newacronym{cn}{CN}{Core Network}
\newacronym{cqi}{CQI}{Channel Quality Information}
\newacronym{cir}{CIR}{Channel Impulse Response}
\newacronym{cp}{CP}{Control Plane}
\newacronym{csirs}{CSI-RS}{Channel State Information - Reference Signal}
\newacronym{dc}{DC}{Dual Connectivity}
\newacronym{dce}{DCE}{Direct Code Execution}
\newacronym{dci}{DCI}{Downlink Control Information}
\newacronym{dl}{DL}{Downlink}
\newacronym{dmr}{DMR}{Deadline Miss Ratio}
\newacronym{dmrs}{DMRS}{DeModulation Reference Signal}
\newacronym{e2e}{E2E}{End-to-End}
\newacronym{ecn}{ECN}{Explicit Congestion Notification}
\newacronym{edf}{EDF}{Earliest Deadline First}
\newacronym{enb}{eNB}{evolved Node Base}
\newacronym{epc}{EPC}{Evolved Packet Core}
\newacronym{es}{ES}{Edge Server}
\newacronym{fdma}{FDMA}{Frequency Division Multiple Access}
\newacronym{fdd}{FDD}{Frequency Division Duplexing}
\newacronym[firstplural=Radio Access Technologies (RATs)]{rat}{RAT}{Radio Access Technology}
\newacronym{fs}{FS}{Fast Switching}
\newacronym{ftp}{FTP}{File Transfer Protocol}
\newacronym{gnb}{gNB}{Next Generation Node Base}
\newacronym{harq}{HARQ}{Hybrid Automatic Repeat reQuest}
\newacronym{hetnet}{HetNet}{Heterogeneous Network}
\newacronym{hh}{HH}{Hard Handover}
\newacronym{hol}{HOL}{Head-of-Line}
\newacronym{ia}{IA}{Initial Access}
\newacronym{imt}{IMT}{International Mobile Telecommunication}
\newacronym{iot}{IoT}{Internet of Things}
\newacronym{lcr}{LCR}{Level Crossing Rate}
\newacronym{lcf}{LCF}{Level Crossing Frequency}
\newacronym{los}{LoS}{Line-of-Sight}
\newacronym{lte}{LTE}{Long Term Evolution}
\newacronym{m2m}{M2M}{Machine to Machine}
\newacronym{mac}{MAC}{Medium Access Control}
\newacronym{mc}{MC}{Multi-Connectivity}
\newacronym{mcs}{MCS}{Modulation and Coding Scheme}
\newacronym{mec}{MEC}{Mobile Edge Cloud}
\newacronym{mi}{MI}{Mutual Information}
\newacronym{mimo}{MIMO}{Multiple Input, Multiple Output}
\newacronym{mmwave}{mmWave}{millimeter wave}
\newacronym{mptcp}{MPTCP}{Multipath TCP}
\newacronym{mr}{MR}{Maximum Rate}
\newacronym{mss}{MSS}{Maximum Segment Size}
\newacronym{mtd}{MTD}{Machine-Type Device}
\newacronym{mtu}{MTU}{Maximum Transmission Unit}
\newacronym{nfv}{NFV}{Network Function Virtualization}
\newacronym{nlos}{NLoS}{Non-Line-of-Sight}
\newacronym{nr}{NR}{New Radio}
\newacronym{o2i}{O2I}{Outdoor-to-Indoor}
\newacronym{ofdm}{OFDM}{Orthogonal Frequency Division Multiplexing}
\newacronym{pdcch}{PDCCH}{Physical Downlonk Control Channel}
\newacronym{pdcp}{PDCP}{Packet Data Convergence Protocol}
\newacronym{pdsch}{PDSCH}{Physical Downlink Shared Channel}
\newacronym{pdu}{PDU}{Packet Data Unit}
\newacronym{pf}{PF}{Proportional Fair}
\newacronym{pgw}{PGW}{Packet Gateway}
\newacronym{phy}{PHY}{Physical}
\newacronym{pbch}{PBCH}{Physical Broadcast Channel}
\newacronym[plural=\gls{mme}s,firstplural=Mobility Management Entities (MMEs)]{mme}{MME}{Mobility Management Entity}
\newacronym{prb}{PRB}{Physical Resource Block}
\newacronym{pss}{PSS}{Primary Synchronization Signal}
\newacronym{pucch}{PUCCH}{Physical Uplink Control Channel}
\newacronym{pusch}{PUSCH}{Physical Uplink Shared Channel}
\newacronym{rach}{RACH}{Random Access Channel}
\newacronym{ran}{RAN}{Radio Access Network}
\newacronym{red}{RED}{Random Early Detection}
\newacronym{rf}{RF}{Radio Frequency}
\newacronym{rlc}{RLC}{Radio Link Control}
\newacronym{rlf}{RLF}{Radio Link Failure}
\newacronym{rrc}{RRC}{Radio Resource Control}
\newacronym{rrm}{RRM}{Radio Resource Management}
\newacronym{rr}{RR}{Round Robin}
\newacronym{rs}{RS}{Remote Server}
\newacronym{rsrp}{RSRP}{Reference Signal Received Power}
\newacronym{rss}{RSS}{Received Signal Strength}
\newacronym{rtt}{RTT}{Round Trip Time}
\newacronym{rw}{RW}{Receive Window}
\newacronym{rx}{RX}{Receiver}
\newacronym{sa}{SA}{standalone}
\newacronym{sack}{SACK}{Selective Acknowledgment}
\newacronym{sap}{SAP}{Service Access Point}
\newacronym{sch}{SCH}{Secondary Cell Handover}
\newacronym{scoot}{SCOOT}{Split Cycle Offset Optimization Technique}
\newacronym{sdma}{SDMA}{Spatial Division Multiple Access}
\newacronym{sinr}{SINR}{Signal-to-Interference-plus-Noise Ratio}
\newacronym{sir}{SIR}{Signal-to-Interference Ratio}
\newacronym{sm}{SM}{Saturation Mode}
\newacronym{snr}{SNR}{Signal-to-Noise Ratio}
\newacronym{son}{SON}{Self-Organizing Network}
\newacronym{ss}{SS}{Synchronization Signal}
\newacronym{srs}{SRS}{Sounding Reference Signal}
\newacronym{sss}{SSS}{Secondary Synchronization Signal}
\newacronym{tb}{TB}{Transport Block}
\newacronym{tcp}{TCP}{Transmission Control Protocol}
\newacronym{tdd}{TDD}{Time Division Duplexing}
\newacronym{tdma}{TDMA}{Time Division Multiple Access}
\newacronym{tfl}{TfL}{Transport for London}
\newacronym{tm}{TM}{Transparent Mode}
\newacronym{trp}{TRP}{Transmitter Receiver Pair}
\newacronym{tti}{TTI}{Transmission Time Interval}
\newacronym{ttt}{TTT}{Time-to-Trigger}
\newacronym{tx}{TX}{Transmitter}
\newacronym{ue}{UE}{User Equipment}
\newacronym{ul}{UL}{Uplink}
\newacronym{uml}{UML}{Unified Modeling Language}
\newacronym{um}{UM}{Unacknowledged Mode}
\newacronym{uma}{UMa}{Urban Macro}
\newacronym{utc}{UTC}{Urban Traffic Control}
\newacronym{vm}{VM}{Virtual Machine}
\newacronym{rsrq}{RSRQ}{Reference Signal Received Quality}
\newacronym{rssi}{RSSI}{Received Signal Strength Indicator}
\newacronym{crs}{CRS}{Cell Reference Signal}
\newacronym{nsa}{NSA}{Non Stand Alone}
\newacronym{mrdc}{MR-DC}{Multi \gls{rat} \gls{dc}}
\newacronym{endc}{EN-DC}{E-UTRAN-\gls{nr} \gls{dc}}
\newacronym{5gc}{5GC}{5G Core}
\newacronym{si}{SI}{Study Item}
\newacronym{iab}{IAB}{Integrated Access and Backhaul}
\newacronym{wf}{WF}{Wired-first}
\newacronym{hqf}{HQF}{Highest-quality-first}
\newacronym{pa}{PA}{Position-aware}
\newacronym{mlr}{MLR}{Maximum-local-rate}
\newacronym{wbf}{WBF}{Wired Bias Function}
\newacronym{mib}{MIB}{Master Information Block}
\newacronym{sib}{SIB}{Secondary Information Block}
\newacronym{kpi}{KPI}{Key Performance Indicator}
\newacronym{ppp}{PPP}{Poisson Point Process}
\newacronym{mpc}{MPC}{Multi Path Component}
\newacronym{rt}{RT}{Ray Tracer}
\newacronym{aoa}{AoA}{Angle of Arrival}
\newacronym{aod}{AoD}{Angle of Departure}
\newacronym{scm}{SCM}{Spatial Channel Model}

%% file: myTikz.tex
\tikzstyle{startstop} = [rectangle, rounded corners, minimum width=2cm, minimum height=0.5cm,text centered, draw=black]
\tikzstyle{io} = [trapezium, trapezium left angle=70, trapezium right angle=110, minimum width=3cm, minimum height=1cm, text centered, draw=black]
\tikzstyle{process} = [rectangle, minimum width=2cm, minimum height=0.5cm, text centered, draw=black, alignb=center]
\tikzstyle{decision} = [ellipse, minimum width=2cm, minimum height=1cm, text centered, draw=black]
\tikzstyle{arrow} = [thick,<->,>=stealth]
\tikzstyle{line} = [thick,>=stealth]
\tikzstyle{darrow} = [thick,<->,>=stealth,dashed]
\tikzstyle{sarrow} = [thick,->,>=stealth]
\tikzstyle{larrow} = [line width=0.1mm,dashdotted,->,>=stealth]

\makeatletter
\def\grd@save@target#1{%
  \def\grd@target{#1}}
\def\grd@save@start#1{%
  \def\grd@start{#1}}
\tikzset{
  grid with coordinates/.style={
    to path={%
      \pgfextra{%
        \edef\grd@@target{(\tikztotarget)}%
        \tikz@scan@one@point\grd@save@target\grd@@target\relax
        \edef\grd@@start{(\tikztostart)}%
        \tikz@scan@one@point\grd@save@start\grd@@start\relax
        \draw[minor help lines] (\tikztostart) grid (\tikztotarget);
        \draw[major help lines] (\tikztostart) grid (\tikztotarget);
        \grd@start
        \pgfmathsetmacro{\grd@xa}{\the\pgf@x/1cm}
        \pgfmathsetmacro{\grd@ya}{\the\pgf@y/1cm}
        \grd@target
        \pgfmathsetmacro{\grd@xb}{\the\pgf@x/1cm}
        \pgfmathsetmacro{\grd@yb}{\the\pgf@y/1cm}
        \pgfmathsetmacro{\grd@xc}{\grd@xa + \pgfkeysvalueof{/tikz/grid with coordinates/major step x}}
        \pgfmathsetmacro{\grd@yc}{\grd@ya + \pgfkeysvalueof{/tikz/grid with coordinates/major step y}}
        \foreach \x in {\grd@xa,\grd@xc,...,\grd@xb}
        \node[anchor=north] at (\x,\grd@ya) {\pgfmathprintnumber{\x}};
        \foreach \y in {\grd@ya,\grd@yc,...,\grd@yb}
        \node[anchor=east] at (\grd@xa,\y) {\pgfmathprintnumber{\y}};
      }
    }
  },
  minor help lines/.style={
    help lines,
    gray,
    line cap =round,
    xstep=\pgfkeysvalueof{/tikz/grid with coordinates/minor step x},
    ystep=\pgfkeysvalueof{/tikz/grid with coordinates/minor step y}
  },
  major help lines/.style={
    help lines,
    line cap =round,
    line width=\pgfkeysvalueof{/tikz/grid with coordinates/major line width},
    xstep=\pgfkeysvalueof{/tikz/grid with coordinates/major step x},
    ystep=\pgfkeysvalueof{/tikz/grid with coordinates/major step y}
  },
  grid with coordinates/.cd,
  minor step x/.initial=.5,
  minor step y/.initial=.2,
  major step x/.initial=1,
  major step y/.initial=1,
  major line width/.initial=1pt,
}
\makeatother

%% file: img/profiling_bars.tex
%
%
\definecolor{mycolor1}{rgb}{0.00000,0.44700,0.74100}%
\definecolor{mycolor2}{rgb}{0.85000,0.32500,0.09800}%
\definecolor{mycolor3}{rgb}{0.92900,0.69400,0.12500}%
\begin{tikzpicture}
\pgfplotsset{every tick label/.append style={font=\scriptsize}}

\begin{axis}[%
width=0.951\fwidth,
height=\fheight,
at={(0\fwidth,0\fheight)},
scale only axis,
xmin=0.7,
xmax=3.3,
xtick={1, 2, 3},
xticklabels={{$2.07$ ms},{$4.74$ ms},{$67.7$ ms}},
ymin=0,
ymax=100,
axis background/.style={fill=white},
xticklabel pos=right,
yticklabel pos=right,
xmajorgrids,
ymajorgrids,
legend style={legend cell align=left, align=left, draw=white!15!black}
]
\end{axis}

\begin{axis}[%
width=0.951\fwidth,
height=\fheight,
at={(0\fwidth,0\fheight)},
scale only axis,
bar width=0.5,
xmin=0.7,
xmax=3.3,
xtick={1,2,3},
xticklabels={{$1 \times 1$},{$1 \times 256$},{$16 \times 4096$}},
xticklabel style={align=center},
xlabel style={font=\footnotesize\color{white!15!black}},
xlabel={Number of antennas elements [RX $\times$ TX]},
ymin=0,
ymax=100,
ylabel style={font=\footnotesize\color{white!15!black}},
ylabel={Time [\%]},
axis background/.style={fill=white},
xmajorgrids,
ymajorgrids,
legend style={at={(0.99,0.01)}, anchor=south east, font=\footnotesize, legend cell align=left, align=left, draw=white!15!black}
]
\addplot[ybar stacked, fill=mycolor1, draw=black, area legend] table[row sep=crcr] {%
1	79.4230769230769\\
2	88.347274300548\\
3	90.3817457442955\\
};
\addplot[forget plot, color=white!15!black] table[row sep=crcr] {%
-0.2	0\\
4.2	0\\
};
\addlegendentry{Computations}

\addplot[ybar stacked, fill=mycolor2, draw=black, area legend] table[row sep=crcr] {%
1	3.77403846153846\\
2	2.06230170175945\\
3	0.276711336472293\\
};
\addplot[forget plot, color=white!15!black] table[row sep=crcr] {%
-0.2	0\\
4.2	0\\
};
\addlegendentry{Random Variables}

\addplot[ybar stacked, fill=mycolor3, draw=black, area legend] table[row sep=crcr] {%
1	16.8028846153846\\
2	9.59042399769253\\
3	9.34154291923216\\
};
\addplot[forget plot, color=white!15!black] table[row sep=crcr] {%
-0.2	0\\
4.2	0\\
};
\addlegendentry{Other}

\end{axis}

\end{tikzpicture}%

%% file: img/chanGenTime_poster.tex
%
%
\definecolor{mycolor1}{rgb}{0.00000,0.44700,0.74100}%
\definecolor{mycolor2}{rgb}{0.85000,0.32500,0.09800}%
\definecolor{mycolor3}{rgb}{0.92900,0.69400,0.12500}%
\definecolor{mycolor4}{rgb}{0.49400,0.18400,0.55600}%
\begin{tikzpicture}
\pgfplotsset{every tick label/.append style={font=\scriptsize}}
\glsunset{gnb}
\glsunset{ue}

\begin{axis}[%
width=0.951\fwidth,
height=\fheight,
at={(0\fwidth,0\fheight)},
scale only axis,
unbounded coords=jump,
xmin=0,
xmax=1050,
xtick={1,64,256,576,1024},
xticklabels={{$1$},{$64$},{$256$},{$576$},{$1024$}},
xlabel style={font=\footnotesize\color{white!15!black}},
xlabel={Number of antenna elements at the \gls{gnb}},
ymin=0,
ymax=50,
ylabel style={font=\footnotesize\color{white!15!black}},
ylabel={Channel Generation Time [ms]},
axis background/.style={fill=white},
xmajorgrids,
ymajorgrids,
legend style={at={(1.25,1.02)}, font=\footnotesize, anchor=south, legend cell align=left, align=left, draw=white!15!black},
legend columns=3
]
\addplot [color=mycolor1, dashed, line width=1.5pt, mark=diamond, mark options={solid, fill=mycolor1, mycolor1}, forget plot]
  table[row sep=crcr]{%
1	8.2815\\
4	8.3455\\
16	8.545\\
64	9.2545\\
256	11.2385\\
576	14.4025\\
1024	18.7525\\
};
\addplot [color=mycolor2, dashed, line width=1.5pt, mark=diamond, mark options={solid, fill=mycolor2, mycolor2}, forget plot]
  table[row sep=crcr]{%
1	1.135\\
4	1.19\\
16	1.2\\
64	1.252\\
256	1.389\\
576	1.601\\
1024	1.892\\
};
\addplot [color=mycolor3, dashed, line width=1.5pt, mark=diamond, mark options={solid, fill=mycolor3, mycolor3}, forget plot]
  table[row sep=crcr]{%
1	5.628\\
4	5.7855\\
16	5.943\\
64	6.4205\\
256	7.7125\\
576	9.872\\
1024	12.793\\
};
\addplot [color=mycolor4, dashed, line width=1.5pt, mark=diamond, mark options={solid, fill=mycolor4, mycolor4}, forget plot]
  table[row sep=crcr]{%
1	0.949\\
4	0.992\\
16	1.0155\\
64	1.052\\
256	1.149\\
576	1.2905\\
1024	1.494\\
};
\addplot [color=mycolor1, line width=1.5pt, mark=*, mark options={solid, fill=mycolor1, mycolor1}, forget plot]
  table[row sep=crcr]{%
1	8.758\\
4	9.0975\\
16	9.577\\
64	11.6395\\
256	18.5065\\
576	29.812\\
1024	45.8635\\
};
\addplot [color=mycolor2, line width=1.5pt, mark=*, mark options={solid, fill=mycolor2, mycolor2}, forget plot]
  table[row sep=crcr]{%
1	1.2115\\
4	1.269\\
16	1.299\\
64	1.531\\
256	2.5765\\
576	4.265\\
1024	6.9355\\
};
\addplot [color=mycolor3, line width=1.5pt, mark=*, mark options={solid, fill=mycolor3, mycolor3}, forget plot]
  table[row sep=crcr]{%
1	6.069\\
4	6.3315\\
16	6.6465\\
64	8.0525\\
256	12.673\\
576	20.23\\
1024	31.096\\
};
\addplot [color=mycolor4, line width=1.5pt, mark=*, mark options={solid, fill=mycolor4, mycolor4}, forget plot]
  table[row sep=crcr]{%
1	1.02\\
4	1.057\\
16	1.09\\
64	1.242\\
256	1.901\\
576	3.192\\
1024	5.128\\
};
\addplot [color=mycolor1, line width=1.5pt]
  table[row sep=crcr]{%
-100	-100\\
};
\addlegendentry{$N=12/20/12, M=20$}

\addplot [color=mycolor2, line width=1.5pt]
  table[row sep=crcr]{%
-100	-100\\
};
\addlegendentry{$N=12/20/12, M=1$}

\addplot [color=mycolor3, line width=1.5pt]
  table[row sep=crcr]{%
-100	-100\\
};
\addlegendentry{$N=8/8/8, M=20$}

\addplot [color=mycolor4, line width=1.5pt]
  table[row sep=crcr]{%
-100	-100\\
};
\addlegendentry{$N=8/8/8, M=1$}

\addplot [color=black, dashed, line width=1.5pt, mark=diamond, mark options={solid, fill=black, black}]
  table[row sep=crcr]{%
-100	-100\\
};
\addlegendentry{num ant \gls{ue}: $1$}

\addplot [color=black, line width=1.5pt, mark=*, mark options={solid, fill=black, black}]
  table[row sep=crcr]{%
-100	-100\\
};
\addlegendentry{num ant \gls{ue}: $16$}

\coordinate (simpl1) at (576, 3.192);
\coordinate (bs1) at (576, 29.812);

\coordinate (simpl2) at (1024, 1.494);
\coordinate (bs2) at (1024,  18.7525);

\end{axis}

\draw[-Triangle,very thick] (simpl1) -- node[xshift=15pt, yshift=15pt] {$9.3 \times$} (bs1);

\draw[-Triangle,very thick] (simpl2) -- node[xshift=-15pt, yshift=0pt] {$12.5 \times$} (bs2);

\end{tikzpicture}%

%% file: img/speedup.tex
%
%
\definecolor{mycolor1}{rgb}{0.00000,0.44700,0.74100}%
\definecolor{mycolor2}{rgb}{0.85000,0.32500,0.09800}%
\definecolor{mycolor3}{rgb}{0.92900,0.69400,0.12500}%
\definecolor{mycolor4}{rgb}{0.49400,0.18400,0.55600}%
\begin{tikzpicture}
\pgfplotsset{every tick label/.append style={font=\scriptsize}}
\begin{axis}[%
width=0.951\fwidth,
height=\fheight,
at={(0\fwidth,0\fheight)},
scale only axis,
unbounded coords=jump,
xmin=0,
xmax=1050,
xtick={1,64,256,576,1024},
xticklabels={{$1$},{$64$},{$256$},{$576$},{$1024$}},
xlabel style={font=\footnotesize\color{white!15!black}},
xlabel={Number of antenna elements at the \gls{gnb}},
ymin=1,
ymax=14,
ytick={1,5,10},
yticklabels={{$1\times$},{$5\times$},{$10\times$}},
ylabel style={font=\footnotesize\color{white!15!black}},
ylabel={Speed-up Factor},
axis background/.style={fill=white},
xmajorgrids,
ymajorgrids,
legend style={at={(0.5,1.01)}, font=\footnotesize, anchor=south, legend cell align=left, align=left, draw=white!15!black},
legend columns=2
]
\addplot [color=mycolor1, dashed, line width=1.5pt, mark=diamond, mark options={solid, fill=mycolor1, mycolor1}, forget plot]
  table[row sep=crcr]{%
1	1\\
4	1\\
16	1\\
64	1\\
256	1\\
576	1\\
1024	1\\
};
\addplot [color=mycolor2, dashed, line width=1.5pt, mark=diamond, mark options={solid, fill=mycolor2, mycolor2}, forget plot]
  table[row sep=crcr]{%
1	7.29647577092511\\
4	7.01302521008403\\
16	7.12083333333333\\
64	7.3917731629393\\
256	8.09107271418286\\
576	8.99594003747658\\
1024	9.91146934460888\\
};
\addplot [color=mycolor3, dashed, line width=1.5pt, mark=diamond, mark options={solid, fill=mycolor3, mycolor3}, forget plot]
  table[row sep=crcr]{%
1	1.47148187633262\\
4	1.44248552415522\\
16	1.43782601379775\\
64	1.44139864496535\\
256	1.45717990275527\\
576	1.45892423014587\\
1024	1.4658406941296\\
};
\addplot [color=mycolor4, dashed, line width=1.5pt, mark=diamond, mark options={solid, fill=mycolor4, mycolor4}, forget plot]
  table[row sep=crcr]{%
1	8.72655426765016\\
4	8.41280241935484\\
16	8.41457410142787\\
64	8.79705323193916\\
256	9.78111401218451\\
576	11.1604029445951\\
1024	12.5518741633199\\
};
\addplot [color=mycolor1, line width=1.5pt, forget plot]
  table[row sep=crcr]{%
-1000	-1000\\
};
\addplot [color=mycolor2, line width=1.5pt, forget plot]
  table[row sep=crcr]{%
-1000	-1000\\
};
\addplot [color=mycolor3, line width=1.5pt, forget plot]
  table[row sep=crcr]{%
-1000	-1000\\
};
\addplot [color=mycolor4, line width=1.5pt, forget plot]
  table[row sep=crcr]{%
-1000	-1000\\
};
\addplot [color=black, dashed, line width=1.5pt, mark=diamond, mark options={solid, fill=black, black}, forget plot]
  table[row sep=crcr]{%
-1000	-1000\\
};
\addplot [color=black, line width=1.5pt, mark=*, mark options={solid, fill=black, black}, forget plot]
  table[row sep=crcr]{%
-1000	-1000\\
};
\addplot [color=mycolor1, line width=1.5pt, mark=*, mark options={solid, fill=mycolor1, mycolor1}, forget plot]
  table[row sep=crcr]{%
1	1\\
4	1\\
16	1\\
64	1\\
256	1\\
576	1\\
1024	1\\
};
\addplot [color=mycolor2, line width=1.5pt, mark=*, mark options={solid, fill=mycolor2, mycolor2}, forget plot]
  table[row sep=crcr]{%
1	7.22905489063145\\
4	7.16903073286052\\
16	7.37259430331024\\
64	7.60254735467015\\
256	7.18280613235009\\
576	6.98991793669402\\
1024	6.61286136543869\\
};
\addplot [color=mycolor3, line width=1.5pt, mark=*, mark options={solid, fill=mycolor3, mycolor3}, forget plot]
  table[row sep=crcr]{%
1	1.44307134618553\\
4	1.43686330253494\\
16	1.44090874896562\\
64	1.44545172306737\\
256	1.4603093190247\\
576	1.47365299060801\\
1024	1.47490030872138\\
};
\addplot [color=mycolor4, line width=1.5pt, mark=*, mark options={solid, fill=mycolor4, mycolor4}, forget plot]
  table[row sep=crcr]{%
1	8.58627450980392\\
4	8.60690633869442\\
16	8.78623853211009\\
64	9.37157809983897\\
256	9.73513940031562\\
576	9.33959899749373\\
1024	8.94374024960998\\
};
\addplot [color=mycolor1, line width=1.5pt]
  table[row sep=crcr]{%
-1000	-1000\\
};

\addplot [color=mycolor2, line width=1.5pt]
  table[row sep=crcr]{%
-1000	-1000\\
};

\addplot [color=mycolor3, line width=1.5pt]
  table[row sep=crcr]{%
-1000	-1000\\
};

\addplot [color=mycolor4, line width=1.5pt]
  table[row sep=crcr]{%
-1000	-1000\\
};

\addplot [color=black, dashed, line width=1.5pt, mark=diamond, mark options={solid, fill=black, black}]
  table[row sep=crcr]{%
-1000	-1000\\
};

\addplot [color=black, line width=1.5pt, mark=*, mark options={solid, fill=black, black}]
  table[row sep=crcr]{%
-1000	-1000\\
};

\end{axis}

\begin{axis}[%
width=1.227\fwidth,
height=1.227\fheight,
at={(-0.16\fwidth,-0.135\fheight)},
scale only axis,
xmin=0,
xmax=1,
ymin=0,
ymax=1,
axis line style={draw=none},
ticks=none,
legend style={legend cell align=left, align=left, draw=white!15!black}
]
\end{axis}
\end{tikzpicture}%

%% file: img/3gpp_8x8_4x4_pdf.tex
%
%
\definecolor{mycolor1}{rgb}{0.00000,0.44700,0.74100}%
\definecolor{mycolor2}{rgb}{0.85000,0.32500,0.09800}%
\definecolor{mycolor3}{rgb}{0.92900,0.69400,0.12500}%
\definecolor{mycolor4}{rgb}{0.49400,0.18400,0.55600}%
\begin{tikzpicture}
\pgfplotsset{every tick label/.append style={font=\scriptsize}}

\begin{axis}[%
width=0.951\fwidth,
height=\fheight,
at={(0\fwidth,0\fheight)},
scale only axis,
xmin=-60,
xmax=60,
xlabel style={font=\footnotesize\color{white!15!black}},
xlabel={Narrowband SINR $\Gamma$ [dB]},
ymin=0,
ymax=0.025,
ylabel shift=-4pt,
yticklabel shift=-1pt,
ylabel style={font=\footnotesize\color{white!15!black}},
ylabel={$p(x)$},
axis background/.style={fill=white},
xmajorgrids,
ymajorgrids,
]
\addplot [color=mycolor1, line width=2.0pt]
  table[row sep=crcr]{%
-60.1033359889037	5.59088949998454e-05\\
-58.6407987942532	8.0081528892606e-05\\
-57.1782615996027	0.00011832675786394\\
-55.7157244049522	0.000172000984953513\\
-52.7906500156511	0.000308165138719119\\
-51.3281128210006	0.00038608323947642\\
-49.8655756263501	0.000484119413144413\\
-48.4030384316996	0.00062526661544382\\
-46.940501237049	0.000832231030010178\\
-45.4779640423985	0.00111427335924219\\
-44.015426847748	0.00146016366151969\\
-42.5528896530975	0.00184913143126408\\
-41.090352458447	0.00227333625526427\\
-39.6278152637964	0.00275021472668868\\
-38.1652780691459	0.0033131840616818\\
-36.7027408744954	0.00398706899632373\\
-35.2402036798449	0.00477349999405163\\
-33.7776664851943	0.00566045558064587\\
-32.3151292905438	0.0066417450616143\\
-30.8525920958933	0.00771545530314199\\
-29.3900549012428	0.00886269417976138\\
-27.9275177065923	0.0100356698842603\\
-26.4649805119417	0.0111704640419319\\
-25.0024433172912	0.0122120512711703\\
-23.5399061226407	0.0131340919205414\\
-22.0773689279902	0.0139422474340094\\
-20.6148317333397	0.0146528083426318\\
-19.1522945386891	0.0152580669348481\\
-17.6897573440386	0.0157279452032952\\
-16.2272201493881	0.0160631943336966\\
-14.7646829547376	0.0163340705400046\\
-13.3021457600871	0.0166390545701702\\
-11.8396085654365	0.0170303489400894\\
-10.377071370786	0.0174966920846558\\
-8.91453417613549	0.0180043724105445\\
-7.45199698148497	0.0185241044345332\\
-5.98945978683444	0.0190172418353853\\
-4.52692259218392	0.0194282593887465\\
-3.0643853975334	0.0197135101929646\\
-1.60184820288289	0.019867705374665\\
-0.139311008232369	0.0198993257862412\\
1.32322618641815	0.0197907017134114\\
2.78576338106868	0.0195190208790024\\
4.2483005757192	0.0191165519551291\\
7.17337496502023	0.0182117158587332\\
8.63591215967077	0.0177501465643601\\
10.0984493543213	0.0172482194355226\\
11.5609865489718	0.0166965640716157\\
13.0235237436223	0.0161036223124071\\
14.4860609382728	0.0154705005679219\\
15.9485981329234	0.0147668739000082\\
17.4111353275739	0.0139331288028259\\
18.8736725222244	0.0129295359078085\\
20.3362097168749	0.0117938612910748\\
21.7987469115254	0.0106317727446879\\
23.261284106176	0.00953212832492767\\
24.7238213008265	0.00849885199023248\\
26.186358495477	0.00747950971954481\\
27.6488956901275	0.00645181834696729\\
29.1114328847781	0.00546294981457862\\
30.5739700794286	0.0045856631395651\\
32.0365072740791	0.00385413373206944\\
33.4990444687296	0.00324768331309144\\
34.9615816633801	0.00272473750380442\\
36.4241188580307	0.00226051419625861\\
37.8866560526812	0.0018550854670849\\
39.3491932473317	0.00151616372411922\\
40.8117304419822	0.0012411826573171\\
42.2742676366327	0.0010144122918021\\
43.7368048312833	0.000817564629905121\\
45.1993420259338	0.00064141940947593\\
46.6618792205843	0.000488582895400214\\
48.1244164152348	0.000367050796093338\\
49.5869536098853	0.000280040325442599\\
51.0494908045359	0.000220059995783117\\
53.9745651938369	0.000129695951059716\\
56.8996395831379	5.46163079349071e-05\\
58.3621767777885	3.13601881600789e-05\\
61.2872511670895	9.83418598110575e-06\\
};

\addplot [color=mycolor2, line width=2.0pt]
  table[row sep=crcr]{%
-61.2395725792646	3.22513973571859e-05\\
-59.8448875245315	4.66886200527483e-05\\
-58.4502024697985	7.00477489843365e-05\\
-57.0555174150654	0.000104452720890436\\
-55.6608323603324	0.00014902567428976\\
-54.2661473055993	0.00020200364176759\\
-52.8714622508663	0.000263886233113908\\
-51.4767771961332	0.000338049551622532\\
-50.0820921414002	0.000427504061050854\\
-48.6874070866672	0.000532690559481352\\
-47.2927220319341	0.00065565503361853\\
-45.8980369772011	0.000808754762722685\\
-44.503351922468	0.00101785736379156\\
-43.108666867735	0.00131457414362757\\
-41.7139818130019	0.00171987765106962\\
-40.3192967582689	0.00223163911299906\\
-38.9246117035358	0.00282909389323294\\
-37.5299266488028	0.00349327438667046\\
-36.1352415940698	0.0042239938278712\\
-34.7405565393367	0.00503915746269001\\
-33.3458714846037	0.00595862649092282\\
-31.9511864298706	0.00698167258419602\\
-30.5565013751376	0.00807256245149546\\
-29.1618163204045	0.0091716720871986\\
-27.7671312656715	0.0102232791753636\\
-26.3724462109384	0.0111947260534819\\
-24.9777611562054	0.0120837131519593\\
-23.5830761014723	0.0129186272362887\\
-22.1883910467393	0.0137378406424915\\
-20.7937059920063	0.0145491260086672\\
-19.3990209372732	0.0153090307619905\\
-18.0043358825402	0.0159490236857778\\
-16.6096508278071	0.0164281990197637\\
-15.2149657730741	0.0167685173425056\\
-13.820280718341	0.0170437301242075\\
-12.425595663608	0.0173297425949457\\
-11.0309106088749	0.0176607102993032\\
-8.24154049940885	0.0183801629629841\\
-5.45217038994276	0.0190581537788859\\
-4.05748533520971	0.0194102257290893\\
-2.66280028047666	0.0197493278445933\\
-1.26811522574363	0.0200088342674363\\
0.126569828989432	0.0201314767015077\\
1.52125488372246	0.0201043300473813\\
2.91593993845551	0.0199502091526327\\
4.31062499318857	0.0197011455167626\\
5.7053100479216	0.0193777799677548\\
7.09999510265465	0.018979090819343\\
8.49468015738771	0.018483293404195\\
9.88936521212074	0.017861700766403\\
11.2840502668538	0.0170996222081072\\
12.6787353215868	0.0162146232670821\\
15.4681054310529	0.0143186210718511\\
16.862790485786	0.0134260515188203\\
19.6521605952521	0.0117399540516629\\
21.0468456499851	0.0108637165658649\\
22.4415307047181	0.00992791881664346\\
23.8362157594512	0.00894105056942607\\
25.2309008141842	0.00794082892728909\\
26.6255858689173	0.00696960575003658\\
28.0202709236503	0.00605064627864493\\
29.4149559783834	0.00519064344020848\\
30.8096410331164	0.00439665320624982\\
32.2043260878495	0.00367901994540887\\
33.5990111425825	0.0030436501009774\\
34.9936961973156	0.00249383729062913\\
36.3883812520486	0.00203764103154214\\
37.7830663067816	0.00168334963412065\\
39.1777513615147	0.00142431057088999\\
40.5724364162477	0.00123414752724926\\
41.9671214709808	0.00108026638478265\\
44.7564915804469	0.000801765563409162\\
46.1511766351799	0.00066751089378414\\
47.545861689913	0.00054114780981962\\
48.940546744646	0.000428554719199781\\
50.335231799379	0.000333079403979752\\
51.7299168541121	0.000253283836343599\\
53.1246019088452	0.000185489660246674\\
54.5192869635782	0.000128639344076475\\
55.9139720183112	8.45618178715313e-05\\
57.3086570730443	5.38279712571921e-05\\
60.0980271825104	2.0126177510349e-05\\
};

\addplot [color=mycolor3, line width=2.0pt]
  table[row sep=crcr]{%
-60.3768599557753	5.02845617376124e-05\\
-58.8920237770161	6.87518700956957e-05\\
-57.407187598257	0.000101271662835245\\
-55.9223514194979	0.000151291931580033\\
-54.4375152407387	0.000220486785671881\\
-52.9526790619796	0.000308244955235182\\
-51.4678428832205	0.000410708707860863\\
-49.9830067044614	0.00052358276293063\\
-48.4981705257022	0.000649151259963787\\
-47.0133343469431	0.000801600738924435\\
-45.528498168184	0.00100730709556274\\
-44.0436619894249	0.00129768583081358\\
-42.5588258106657	0.00169246625299024\\
-41.0739896319066	0.00218461145536253\\
-39.5891534531475	0.00275100599984057\\
-38.1043172743884	0.00338557473457257\\
-36.6194810956292	0.00411286704463265\\
-35.1346449168701	0.00495936263340013\\
-33.649808738111	0.00592053637654288\\
-32.1649725593519	0.00696422706676714\\
-30.6801363805927	0.00805890075896798\\
-29.1953002018336	0.00918476077325892\\
-27.7104640230745	0.0103187940654266\\
-26.2256278443153	0.0114199300498541\\
-24.7407916655562	0.0124360506348609\\
-23.2559554867971	0.0133241258808923\\
-21.771119308038	0.014059211452718\\
-20.2862831292788	0.01462835723148\\
-18.8014469505197	0.0150374886151141\\
-17.3166107717606	0.0153331887433126\\
-15.8317745930015	0.0156017092375649\\
-14.3469384142423	0.0159235290640964\\
-12.8621022354832	0.0163347930879354\\
-11.3772660567241	0.016834463002148\\
-9.89242987796496	0.0174027556435021\\
-8.40759369920583	0.0179965183824535\\
-6.92275752044671	0.0185600652548317\\
-5.43792134168758	0.0190521467674074\\
-3.95308516292846	0.0194396283546183\\
-2.46824898416934	0.019676190116293\\
-0.983412805410211	0.0197349676956406\\
0.501423373348928	0.0196585388248636\\
1.98625955210805	0.019528667557104\\
3.47109573086718	0.0193820388661905\\
4.9559319096263	0.0191802670980792\\
6.44076808838544	0.0188605209646937\\
7.92560426714456	0.0183903942473833\\
9.41044044590369	0.0177707677168542\\
10.8952766246628	0.0170185362118147\\
12.3801128034219	0.0161735727011987\\
13.8649489821811	0.0153029955205639\\
15.3497851609402	0.0144609755299356\\
16.8346213396993	0.0136432876002317\\
18.3194575184584	0.01279881803611\\
19.8042936972176	0.0118905049503226\\
21.2891298759767	0.0109427268231741\\
22.7739660547358	0.0100225285715254\\
24.2588022334949	0.00916818505447736\\
25.7436384122541	0.00833916930460532\\
27.2284745910132	0.00745053055155864\\
28.7133107697723	0.00646067268330341\\
30.1981469485314	0.00542004902340665\\
31.6829831272906	0.00443622478285732\\
33.1678193060497	0.00359858596750229\\
34.6526554848088	0.00293471949902369\\
36.137491663568	0.00242231172602203\\
37.6223278423271	0.00202179954874282\\
39.1071640210862	0.00169449766920593\\
40.5920001998453	0.00140990198778468\\
42.0768363786045	0.00115322533635975\\
43.5616725573636	0.000926375781432398\\
45.0465087361227	0.000735789444973989\\
46.5313449148818	0.000579540784571009\\
48.016181093641	0.000448973621011817\\
49.5010172724001	0.000338872502751997\\
50.9858534511592	0.000249632172319991\\
52.4706896299183	0.000180496787521633\\
53.9555258086775	0.000126391659193814\\
55.4403619874366	8.26851402209172e-05\\
56.9251981661957	4.91188907005835e-05\\
58.4100343449549	2.70421273285137e-05\\
61.3797067024731	9.9872374903498e-06\\
};

\addplot [color=mycolor4, line width=2.0pt]
  table[row sep=crcr]{%
-60.7591562058255	6.53127109373486e-05\\
-57.9175344258895	0.000133909378888575\\
-55.0759126459534	0.000207566984073537\\
-52.2342908660173	0.000272882054503043\\
-50.8134799760493	0.000328971689732782\\
-49.3926690860813	0.000422965559963018\\
-47.9718581961132	0.000565619395963779\\
-46.5510473061452	0.000756988386825697\\
-45.1302364161772	0.000989634655830685\\
-43.7094255262092	0.00126129094442007\\
-42.2886146362411	0.00158659947002349\\
-40.8678037462731	0.00199179509141345\\
-39.4469928563051	0.00249752728430508\\
-38.026181966337	0.00311133256573726\\
-36.605371076369	0.00383760615972051\\
-35.184560186401	0.00468607139956845\\
-33.7637492964329	0.00566307875937611\\
-32.3429384064649	0.006759224690704\\
-30.9221275164969	0.00794284355474417\\
-29.5013166265289	0.0091567268791124\\
-28.0805057365608	0.0103252933291316\\
-26.6596948465928	0.0113796158250068\\
-25.2388839566248	0.01228708851108\\
-23.8180730666567	0.0130524491512247\\
-22.3972621766887	0.0136917775735554\\
-20.9764512867207	0.0142161785918802\\
-19.5556403967527	0.0146529645451139\\
-18.1348295067846	0.0150626043488202\\
-16.7140186168166	0.0155020324854789\\
-15.2932077268486	0.0159714461819291\\
-13.8723968368805	0.0164226320740468\\
-12.4515859469125	0.0168253391307189\\
-11.0307750569445	0.0172074866024658\\
-9.60996416697645	0.0176272357233032\\
-8.18915327700842	0.0181161286789759\\
-6.76834238704039	0.0186506857107958\\
-5.34753149707235	0.0191689941981394\\
-3.92672060710433	0.0196084754432562\\
-2.50590971713629	0.0199291132423056\\
-1.08509882716827	0.0201220208879747\\
0.335712062799772	0.0202086968887372\\
1.7565229527678	0.0202180057646615\\
3.17733384273583	0.0201381780048493\\
4.59814473270386	0.0199084522461845\\
6.01895562267188	0.019476431549748\\
7.43976651263991	0.0188491014211181\\
8.86057740260794	0.018078605094253\\
10.281388292576	0.017220280104759\\
13.123010072512	0.0154161394127286\\
14.5438209624801	0.0145308836705738\\
15.9646318524481	0.0136748795118606\\
17.3854427424161	0.0128586285738805\\
18.8062536323841	0.0120952397023331\\
20.2270645223522	0.0113826017710821\\
21.6478754123202	0.0106914990492299\\
23.0686863022882	0.00997598331213823\\
24.4894971922563	0.00919090509846399\\
25.9103080822243	0.0083037892316753\\
27.3311189721923	0.00731001040588808\\
28.7519298621604	0.00625074443397722\\
30.1727407521284	0.00520850920862159\\
31.5935516420964	0.00427238911672845\\
33.0143625320644	0.00349791242545194\\
34.4351734220325	0.002889392211884\\
35.8559843120005	0.0024145026726714\\
37.2767952019685	0.0020327655330874\\
38.6976060919366	0.00171197138429591\\
40.1184169819046	0.00142939241973039\\
41.5392278718726	0.00117092932364926\\
42.9600387618407	0.000935344528102178\\
44.3808496518087	0.000733930498277857\\
45.8016605417767	0.000578995968673723\\
47.2224714317448	0.000469767911582153\\
48.6432823217128	0.000390416455587683\\
51.4849041016488	0.000257695195010399\\
52.9057149916169	0.000198106253428421\\
54.3265258815849	0.000148976384558352\\
55.7473367715529	0.000111466166828222\\
58.588958551489	5.74905825274641e-05\\
60.009769441457	3.62061074596909e-05\\
};

\end{axis}

\end{tikzpicture}%

%% file: img/lcr_afd_manyAnt_svd_AFD.tex
%
%
\definecolor{mycolor1}{rgb}{0.92900,0.69400,0.12500}%
\definecolor{mycolor2}{rgb}{0.00000,0.44700,0.74100}%
\definecolor{mycolor3}{rgb}{0.85000,0.32500,0.09800}%
\definecolor{mycolor4}{rgb}{0.49400,0.18400,0.55600}%
\begin{tikzpicture}
\pgfplotsset{every tick label/.append style={font=\scriptsize}}

\begin{axis}[%
width=0.951\fwidth,
height=\fheight,
at={(0\fwidth,0\fheight)},
scale only axis,
xmin=-40,
xmax=60,
xtick={-40,-20,0,20,40,60},
ymode=log,
ymin=0.375366210937498,
ymax=1000000,
ytick={    1,   100, 10000},
yminorticks=true,
ylabel style={font=\footnotesize\color{white!15!black}, align=center},
xlabel style={font=\footnotesize\color{white!15!black}, align=center},
ylabel={AFBW [kHz]},
xlabel={Wideband SIR threshold $\xi_{th}$ [dB]},
axis background/.style={fill=white},
xmajorgrids,
ymajorgrids,
yminorgrids,
legend style={font=\footnotesize, at={(0.5, 1.01)}, anchor=south, legend cell align=left, align=left, draw=white!15!black},
legend columns=2,
]
\addplot [color=mycolor1, line width=1.5pt]
  table[row sep=crcr]{%
-40.1201201201201	0.547281901041669\\
-39.95995995996	0.582275390624998\\
-39.7997997997998	0.601806640624999\\
-39.6396396396396	0.613199869791664\\
-39.4794794794795	0.630289713541667\\
-39.3193193193193	0.659586588541671\\
-38.998998998999	0.694986979166671\\
-38.6786786786787	0.714111328125005\\
-38.3583583583584	0.770263671875\\
-38.038038038038	0.807698567708333\\
-37.7177177177177	0.881754557291665\\
-37.5575575575576	0.930582682291671\\
-37.2372372372372	0.996704101562505\\
-36.7567567567568	1.08723958333334\\
-36.5965965965966	1.10921223958333\\
-36.4364364364364	1.15275065104166\\
-36.2762762762763	1.18184407552082\\
-36.1161161161161	1.23441569010416\\
-35.955955955956	1.2646484375\\
-35.6356356356356	1.34916178385416\\
-35.4754754754755	1.40653483072916\\
-34.6746746746747	1.63114420572916\\
-34.5145145145145	1.70397949218749\\
-34.3543543543544	1.76468912760415\\
-33.8738738738739	1.88968912760417\\
-33.7137137137137	1.96964518229166\\
-33.5535535535536	2.01837158203126\\
-33.2332332332332	2.17795817057291\\
-32.5925925925926	2.4855007595486\\
-32.4324324324324	2.5429958767361\\
-32.2722722722723	2.64048936631944\\
-32.1121121121121	2.70254177517359\\
-31.951951951952	2.80862087673611\\
-31.7917917917918	2.88837348090278\\
-31.6316316316316	2.99961441282242\\
-31.1511511511512	3.26420665922619\\
-30.1901901901902	3.98657731691756\\
-29.8698698698699	4.25615424614446\\
-29.7097097097097	4.35392966002094\\
-29.2292292292292	4.87035724891653\\
-29.0690690690691	4.97557826152799\\
-28.9089089089089	5.15621400490782\\
-28.2682682682683	5.79120712458114\\
-28.1081081081081	6.02153847942422\\
-27.9479479479479	6.18851758796214\\
-27.7877877877878	6.42091462614535\\
-26.986986986987	7.57776579093108\\
-26.6666666666667	7.966745477152\\
-26.3463463463463	8.47646213988377\\
-26.026026026026	8.90988329851428\\
-25.5455455455455	9.8578810038161\\
-24.4244244244244	12.1212578382988\\
-24.2642642642643	12.5572063208783\\
-24.1041041041041	12.8692386782598\\
-23.6236236236236	14.0945461256447\\
-23.3033033033033	14.7803015755552\\
-22.6626626626627	16.756659908653\\
-22.3423423423423	17.597166703801\\
-20.1001001001001	26.1566825153199\\
-19.9399399399399	26.6354130578854\\
-19.2992992992993	29.718625500011\\
-18.8188188188188	32.4747164245386\\
-16.5765765765766	47.3595451496529\\
-16.0960960960961	50.3546879062243\\
-15.7757757757758	52.8143794547788\\
-14.4944944944945	65.2690070042655\\
-13.6936936936937	73.3241793914735\\
-12.8928928928929	83.3702739627232\\
-10.8108108108108	111.808499867353\\
-10.3303303303303	118.243794719608\\
-10.1701701701702	122.047972961497\\
-9.36936936936937	136.945760931047\\
-9.20920920920921	142.977691428029\\
-9.04904904904905	146.18705410978\\
-8.72872872872873	151.787458790909\\
-6.8068068068068	192.821795808855\\
-6.64664664664664	195.206158604878\\
-6.32632632632632	203.558802047131\\
-5.52552552552552	220.656128774746\\
-5.04504504504504	236.840794938827\\
-3.76376376376376	271.348233513355\\
-1.52152152152152	355.920637573825\\
-1.2012012012012	367.786990086669\\
-1.04104104104104	377.609953761781\\
-0.880880880880881	384.39425960858\\
-0.72072072072072	395.733586246367\\
-0.56056056056056	401.076274338572\\
0.0800800800800801	433.793676627973\\
1.68168168168168	530.474801122663\\
1.84184184184184	537.395676404544\\
2.16216216216216	561.127057869674\\
2.8028028028028	604.532333489563\\
3.6036036036036	680.267198837671\\
4.08408408408408	715.867312315913\\
4.72472472472472	792.411358377425\\
5.2052052052052	838.164350648356\\
5.36536536536536	862.684619234576\\
5.68568568568568	891.727744141121\\
6.16616616616616	956.368844006263\\
6.96696696696696	1058.51348816088\\
7.6076076076076	1123.59834539391\\
8.40840840840841	1236.07251514897\\
8.72872872872873	1301.11413830669\\
9.20920920920921	1385.73557189386\\
11.1311311311311	1763.70076738652\\
12.0920920920921	2014.02408569946\\
12.5725725725726	2130.76940270622\\
12.8928928928929	2230.30764707403\\
13.053053053053	2265.647280812\\
13.3733733733734	2369.64089740365\\
15.6156156156156	3167.61724853155\\
17.2172172172172	3806.99688999971\\
19.2992992992993	4795.04799672989\\
19.9399399399399	5147.10620423203\\
21.3813813813814	6070.94829450909\\
23.3033033033033	7447.66912880011\\
26.8268268268268	10569.0920509052\\
30.03003003003	14342.3620901888\\
34.034034034034	20685.4107721394\\
36.5965965965966	25682.1775267295\\
39.95995995996	33731.1577518406\\
42.8428428428428	41474.673204241\\
46.5265265265265	52093.8066905847\\
51.1711711711712	65705.4326374729\\
52.1321321321321	68314.0300628108\\
54.2142142142142	73665.0901949041\\
58.3783783783784	82865.9043380024\\
60.1401401401401	85993.068756119\\
};
\addlegendentry{$N=8/8/8, M=20$}

\addplot [color=mycolor2, line width=1.5pt]
  table[row sep=crcr]{%
-40.1201201201201	0.622558593750002\\
-39.95995995996	0.639648437499995\\
-39.7997997997998	0.664062500000002\\
-39.4794794794795	0.725097656249994\\
-38.6786786786787	0.860595703125\\
-38.5185185185185	0.906982421875\\
-38.1981981981982	0.981445312500008\\
-37.7177177177177	1.083984375\\
-37.5575575575576	1.12792968750001\\
-36.9169169169169	1.280517578125\\
-36.5965965965966	1.39038085937499\\
-36.4364364364364	1.43025716145834\\
-36.2762762762763	1.48885091145833\\
-35.955955955956	1.58406575520834\\
-35.7957957957958	1.64510091145833\\
-35.6356356356356	1.69209798177084\\
-35.4754754754755	1.75679524739584\\
-34.6746746746747	2.02880859375\\
-34.034034034034	2.33426920572916\\
-33.5535535535536	2.55987839471728\\
-33.3933933933934	2.66888718377974\\
-32.7527527527528	2.97733561197918\\
-32.2722722722723	3.30546061197918\\
-32.1121121121121	3.39355468749998\\
-31.4714714714715	3.90302966889883\\
-30.1901901901902	5.09416694145701\\
-30.03003003003	5.229423754143\\
-29.2292292292292	6.17017443683707\\
-28.9089089089089	6.66741543639736\\
-28.7487487487487	6.80271509699605\\
-28.4284284284284	7.28245159110633\\
-28.2682682682683	7.46428113444716\\
-28.1081081081081	7.72336110476976\\
-27.9479479479479	7.94580857578298\\
-27.7877877877878	8.23495003396133\\
-27.1471471471471	9.10839869235612\\
-26.8268268268268	9.72098493352189\\
-26.6666666666667	9.93848984172041\\
-25.8658658658659	11.4637997094052\\
-20.1001001001001	30.0612716171915\\
-19.2992992992993	33.7171897224668\\
-16.7367367367367	51.2982773971326\\
-16.5765765765766	52.3391182759275\\
-16.2562562562563	55.1881333018364\\
-13.3733733733734	84.3650418734034\\
-13.2132132132132	85.8114862898391\\
-10.1701701701702	130.141569933814\\
-8.88888888888889	153.388588292683\\
-2.96296296296296	305.031182521932\\
-2.8028028028028	312.221375252996\\
-1.2012012012012	368.45012044962\\
-1.04104104104104	379.166492603311\\
-0.56056056056056	403.409868376841\\
-0.0800800800800801	419.651920021077\\
1.04104104104104	471.07414461816\\
1.36136136136136	483.483564098835\\
2.002002002002	518.459721160664\\
2.16216216216216	533.811118915344\\
2.64264264264264	561.076000565103\\
2.8028028028028	576.713942643045\\
2.96296296296296	588.119584657278\\
3.12312312312312	604.95314510559\\
3.6036036036036	644.706494505454\\
4.4044044044044	704.675806512621\\
4.56456456456456	723.808402719091\\
6.006006006006	850.317368845052\\
6.16616616616616	872.809224377059\\
6.8068068068068	932.946403944198\\
6.96696696696696	960.522117965474\\
7.12712712712712	974.228689188489\\
7.28728728728728	999.220630135528\\
7.6076076076076	1035.04403294038\\
7.76776776776777	1061.17712652494\\
8.08808808808809	1091.07463114821\\
8.24824824824825	1120.51382357468\\
10.970970970971	1551.32276847864\\
11.9319319319319	1765.33870596693\\
15.6156156156156	2731.37507910443\\
16.2562562562563	2927.58015358402\\
16.5765765765766	3057.2555303986\\
20.1001001001001	4629.26858373901\\
20.4204204204204	4821.78403660771\\
22.8228228228228	6322.26119191814\\
24.2642642642643	7390.6452452507\\
33.2332332332332	18512.9445115066\\
37.7177177177177	28244.7083097597\\
39.95995995996	34245.6022853466\\
41.7217217217217	39420.789510218\\
44.2842842842843	46867.5487696029\\
47.4874874874875	56803.2446341227\\
48.4484484484485	59741.3833321293\\
50.3703703703704	65398.3117877343\\
53.4134134134134	73570.4970598822\\
55.6556556556557	78818.2521680032\\
57.4174174174174	82494.8826465375\\
60.1401401401401	87403.8332134089\\
};
\addlegendentry{$N=12/20/12, M=20$}

\addplot [color=mycolor3, line width=1.5pt]
  table[row sep=crcr]{%
-40.1201201201201	0.544026692708335\\
-39.95995995996	0.561930338541668\\
-39.7997997997998	0.589599609375\\
-39.6396396396396	0.599365234374999\\
-39.3193193193193	0.655517578124998\\
-39.1591591591592	0.678710937499996\\
-38.8388388388388	0.739746093750003\\
-38.5185185185185	0.780029296875004\\
-38.3583583583584	0.813395182291663\\
-38.1981981981982	0.833333333333333\\
-38.038038038038	0.867106119791665\\
-37.8778778778779	0.913492838541672\\
-37.5575575575576	0.996907552083331\\
-37.3973973973974	1.02132161458333\\
-37.2372372372372	1.03902180989583\\
-37.0770770770771	1.06319173177083\\
-36.9169169169169	1.11446126302083\\
-36.5965965965966	1.18159993489584\\
-36.4364364364364	1.22880045572916\\
-36.2762762762763	1.26957194010416\\
-36.1161161161161	1.30057779947916\\
-35.7957957957958	1.41028994605656\\
-35.6356356356356	1.44202822730654\\
-35.3153153153153	1.58993675595238\\
-35.1551551551552	1.64202008928571\\
-34.994994994995	1.70753115699404\\
-34.8348348348348	1.76470075334821\\
-34.5145145145145	1.90606756524724\\
-34.1941941941942	1.99680649753891\\
-33.7137137137137	2.17650204613094\\
-33.5535535535536	2.26944986979165\\
-33.3933933933934	2.34100632440477\\
-33.2332332332332	2.45317150297621\\
-32.5925925925926	2.75357200985862\\
-32.1121121121121	3.05673459976441\\
-31.4714714714715	3.44078524150545\\
-31.3113113113113	3.49984402126738\\
-30.990990990991	3.69508192274304\\
-30.6706706706707	3.94453260633679\\
-30.3503503503504	4.1806902204241\\
-30.1901901901902	4.26687141390513\\
-29.7097097097097	4.6346261160714\\
-29.3893893893894	4.9487429311953\\
-29.2292292292292	5.0686324478744\\
-29.0690690690691	5.22287846024419\\
-28.9089089089089	5.4152520249932\\
-27.3073073073073	7.23141120083263\\
-26.026026026026	8.77411845488308\\
-25.7057057057057	9.32374789042879\\
-25.2252252252252	10.0452210639849\\
-24.9049049049049	10.6291956362577\\
-24.1041041041041	12.0277040048759\\
-23.1431431431431	14.0752768329468\\
-19.1391391391391	25.5416587234618\\
-18.8188188188188	26.5493658498261\\
-16.4164164164164	37.5882671836427\\
-10.970970970971	76.8655278893814\\
-10.8108108108108	78.7087200158316\\
-10.1701701701702	84.9333129271943\\
-9.36936936936937	94.1531239481904\\
-8.24824824824825	107.077713982927\\
-7.28728728728728	120.141760818577\\
-6.96696696696696	124.932678421457\\
-3.6036036036036	185.663473720062\\
-3.44344344344344	192.113029888698\\
-3.28328328328328	191.263727371644\\
-3.12312312312312	199.997964702553\\
-2.96296296296296	201.756700661486\\
-2.8028028028028	209.36806161159\\
-1.52152152152152	238.205561358381\\
-0.880880880880881	255.33712232498\\
-0.72072072072072	263.004997551491\\
0.4004004004004	296.955786492541\\
0.56056056056056	305.61188836687\\
1.2012012012012	325.210980981069\\
1.52152152152152	339.436304316985\\
2.002002002002	359.773963151428\\
2.48248248248248	388.015261078191\\
2.96296296296296	413.316271964402\\
3.12312312312312	426.564980977378\\
6.006006006006	675.685847666318\\
6.16616616616616	688.46003626566\\
6.64664664664664	756.663934047866\\
6.8068068068068	788.755236981585\\
7.6076076076076	892.691180988337\\
7.76776776776777	922.078330495094\\
7.92792792792793	938.888357817288\\
9.84984984984985	1326.2522566414\\
12.5725725725726	2044.14014569908\\
13.5335335335335	2335.13649271974\\
13.8538538538538	2461.13979034565\\
15.2952952952953	2983.54239472311\\
15.6156156156156	3121.55171804375\\
18.6586586586587	4576.8205163316\\
20.7407407407407	5841.27794052475\\
22.982982982983	7511.43016177708\\
23.7837837837838	8127.5640083467\\
28.5885885885886	12884.4098475137\\
35.4754754754755	23774.2509011946\\
40.4404404404404	35490.2119360152\\
41.0810810810811	37151.9103479923\\
44.2842842842843	46411.0216817143\\
46.6866866866867	53638.4539566504\\
47.967967967968	57420.3774159381\\
49.4094094094094	61531.6509546635\\
52.9329329329329	71196.6178612431\\
55.1751751751752	76725.1277171679\\
57.4174174174174	81566.308137376\\
60.1401401401401	86447.6997130776\\
};
\addlegendentry{$N=12/20/12, M=1$}

\addplot [color=mycolor4, line width=1.5pt]
  table[row sep=crcr]{%
-40.1201201201201	0.3662109375\\
-39.95995995996	0.378417968749998\\
-39.7997997997998	0.396728515625003\\
-39.6396396396396	0.401611328125\\
-39.4794794794795	0.416259765625002\\
-39.1591591591592	0.457763671874999\\
-38.998998998999	0.472412109375002\\
-38.8388388388388	0.479736328124997\\
-38.3583583583584	0.522460937500001\\
-37.7177177177177	0.599365234374999\\
-37.2372372372372	0.649414062500004\\
-37.0770770770771	0.675048828125005\\
-36.9169169169169	0.69091796875\\
-36.7567567567568	0.725097656249994\\
-36.4364364364364	0.761311848958336\\
-36.2762762762763	0.766194661458337\\
-36.1161161161161	0.789388020833339\\
-35.955955955956	0.822143554687506\\
-35.7957957957958	0.841674804687505\\
-35.4754754754755	0.896606445312495\\
-35.3153153153153	0.937906901041664\\
-34.994994994995	0.983072916666672\\
-34.5145145145145	1.09436035156251\\
-34.3543543543544	1.14644368489583\\
-34.1941941941942	1.17879231770834\\
-34.034034034034	1.22334798177084\\
-33.8738738738739	1.25345865885416\\
-33.7137137137137	1.30187988281249\\
-33.5535535535536	1.3421630859375\\
-33.3933933933934	1.39587402343751\\
-33.2332332332332	1.44173177083334\\
-33.0730730730731	1.49906412760417\\
-32.9129129129129	1.56905110677084\\
-32.5925925925926	1.65030924479167\\
-32.2722722722723	1.76383463541667\\
-32.1121121121121	1.80187988281249\\
-31.951951951952	1.87707519531249\\
-31.7917917917918	1.92875162760417\\
-31.6316316316316	2.01053873697918\\
-31.4714714714715	2.05743698846725\\
-31.3113113113113	2.13643391927083\\
-31.1511511511512	2.16278076171875\\
-30.990990990991	2.22361246744793\\
-30.8308308308308	2.30885823567707\\
-30.6706706706707	2.37528483072917\\
-30.3503503503504	2.56193033854165\\
-29.8698698698699	2.81519426618304\\
-29.7097097097097	2.88174583798363\\
-29.5495495495495	2.97500755673365\\
-28.5885885885886	3.48015194847472\\
-27.7877877877878	3.99321884441514\\
-26.5065065065065	5.0584099399407\\
-23.1431431431431	8.71957233255763\\
-22.982982982983	8.99377880327603\\
-22.6626626626627	9.44104140452163\\
-22.022022022022	10.5042161171003\\
-21.5415415415415	11.2093880963391\\
-20.9009009009009	12.3925113617868\\
-20.7407407407407	12.6037336913255\\
-20.5805805805806	13.030322687598\\
-18.4984984984985	17.8139509994729\\
-17.5375375375375	20.3185405376252\\
-17.3773773773774	20.9739637533886\\
-16.8968968968969	22.4069962702632\\
-16.0960960960961	25.3763430862481\\
-15.6156156156156	27.1300349454365\\
-15.4554554554554	27.9149136975025\\
-12.0920920920921	44.9025329624662\\
-11.6116116116116	47.4205688305512\\
-10.970970970971	51.4754890623307\\
-10.8108108108108	52.028909184815\\
-10.4904904904905	54.9493402580113\\
-9.84984984984985	59.5223832293277\\
-8.88888888888889	67.5284071087841\\
-7.44744744744744	81.4893498783902\\
-6.8068068068068	89.8586952487317\\
-6.64664664664664	91.1372589099354\\
-5.36536536536536	108.053066408855\\
-5.04504504504504	113.401254407086\\
-4.72472472472472	117.272645472799\\
-4.56456456456456	121.043196412947\\
-4.4044044044044	126.031284848734\\
-2.96296296296296	148.069344946065\\
-2.64264264264264	155.771885171178\\
-2.48248248248248	161.971949174926\\
-2.16216216216216	168.174899695546\\
-1.68168168168168	181.967353603379\\
-1.52152152152152	190.58887439595\\
-1.2012012012012	199.694867689645\\
-1.04104104104104	206.506321178894\\
-0.56056056056056	217.081244712266\\
0.24024024024024	246.710028435753\\
0.4004004004004	254.854583334929\\
0.56056056056056	260.576289797192\\
0.72072072072072	268.438131242371\\
0.880880880880881	273.075371822678\\
1.04104104104104	282.270670652826\\
1.2012012012012	288.01381863019\\
1.36136136136136	301.052943995305\\
1.52152152152152	312.364917815584\\
2.002002002002	335.214962315319\\
2.16216216216216	352.261699876844\\
2.32232232232232	358.438751093185\\
2.48248248248248	372.613044833287\\
2.64264264264264	381.210348003092\\
2.8028028028028	393.657253779222\\
3.12312312312312	415.322140626035\\
4.08408408408408	498.587102580179\\
4.4044044044044	544.834695555598\\
4.72472472472472	581.321187642081\\
5.36536536536536	653.833953338549\\
5.68568568568568	698.793645957989\\
5.84584584584584	740.195889348478\\
6.32632632632632	808.188963433183\\
6.64664664664664	875.160814088537\\
6.8068068068068	905.541382978069\\
6.96696696696696	926.047348471233\\
7.44744744744744	1008.20992068286\\
7.92792792792793	1131.77216968231\\
8.56856856856857	1268.58809729367\\
9.36936936936937	1486.04228852375\\
10.4904904904905	1829.13192119977\\
10.6506506506506	1916.20145067992\\
10.8108108108108	1973.25415843143\\
11.4514514514514	2177.56550290947\\
12.2522522522522	2511.3831624948\\
13.2132132132132	2938.57039190865\\
13.3733733733734	2986.38805803203\\
13.8538538538538	3210.08191084091\\
14.1741741741742	3343.76007684018\\
14.6546546546546	3529.79109500517\\
15.2952952952953	3853.60118443516\\
15.9359359359359	4169.91094935404\\
16.7367367367367	4683.18422859731\\
19.9399399399399	6886.19961789615\\
22.6626626626627	9201.1447396234\\
24.1041041041041	10542.3713001975\\
26.1861861861862	12635.7351414465\\
29.8698698698699	17066.9097404076\\
33.0730730730731	21837.5931570368\\
34.8348348348348	24779.0148049931\\
40.1201201201201	35603.2635937091\\
41.7217217217217	39452.7814341091\\
45.8858858858859	50342.5419004575\\
48.7687687687688	58439.4790660294\\
51.1711711711712	65026.6619436169\\
53.4134134134134	70930.7096901514\\
54.6946946946947	74023.2748526325\\
58.3783783783784	82196.5854278158\\
60.1401401401401	85404.7545159775\\
};
\addlegendentry{$N=8/8/8, M=1$}

\end{axis}
\end{tikzpicture}%

%% file: img/lcr_afd_manyAnt_svd_LCR.tex
%
%
\definecolor{mycolor1}{rgb}{0.00000,0.44700,0.74100}%
\definecolor{mycolor2}{rgb}{0.85000,0.32500,0.09800}%
\definecolor{mycolor3}{rgb}{0.92900,0.69400,0.12500}%
\definecolor{mycolor4}{rgb}{0.49400,0.18400,0.55600}%
\begin{tikzpicture}
\pgfplotsset{every tick label/.append style={font=\scriptsize}}

\begin{axis}[%
width=0.951\fwidth,
height=\fheight,
at={(0\fwidth,0\fheight)},
scale only axis,
xmin=-40,
xmax=60,
xlabel style={font=\color{white!15!black}},
xlabel={Wideband SIR threshold $\xi_{th}$ [dB]},
ymode=log,
ymin=1e-06,
ymax=0.1,
ytick={ 1e-05, 0.0001,  0.001,   0.01},
yminorticks=true,
ylabel style={font=\footnotesize\color{white!15!black}, align=center},
xlabel style={font=\footnotesize\color{white!15!black}, align=center},
ylabel={LCF},
axis background/.style={fill=white},
xmajorgrids,
ymajorgrids,
yminorgrids,
legend style={at={(0.86,0.01)}, font=\footnotesize, anchor=south east, legend cell align=left, align=left, draw=white!15!black},
]
\addplot [color=mycolor1, line width=1.5pt]
  table[row sep=crcr]{%
-40.1201201201201	5.66406249999996e-06\\
-39.95995995996	5.83496093749998e-06\\
-39.7997997997998	6.05468749999997e-06\\
-39.6396396396396	6.32324218749999e-06\\
-39.4794794794795	6.66503906249994e-06\\
-38.998998998999	7.39746093750003e-06\\
-38.8388388388388	7.61718749999993e-06\\
-38.6786786786787	7.88574218749996e-06\\
-38.5185185185185	8.34960937499992e-06\\
-38.3583583583584	8.59374999999993e-06\\
-38.1981981981982	8.93554687499992e-06\\
-37.8778778778779	9.54589843750003e-06\\
-37.7177177177177	9.96093750000004e-06\\
-37.5575575575576	1.05712890625e-05\\
-37.3973973973974	1.08886718750001e-05\\
-37.0770770770771	1.17431640625001e-05\\
-36.9169169169169	1.20361328125e-05\\
-36.5965965965966	1.31347656250001e-05\\
-36.4364364364364	1.35009765624999e-05\\
-36.2762762762763	1.40625e-05\\
-36.1161161161161	1.45263671875e-05\\
-35.955955955956	1.5087890625e-05\\
-35.6356356356356	1.59912109374999e-05\\
-35.4754754754755	1.66259765625e-05\\
-35.3153153153153	1.71142578125001e-05\\
-35.1551551551552	1.77978515625e-05\\
-34.994994994995	1.82617187499999e-05\\
-34.8348348348348	1.90673828125e-05\\
-34.6746746746747	1.97509765624998e-05\\
-34.5145145145145	2.05566406249999e-05\\
-34.1941941941942	2.197265625e-05\\
-34.034034034034	2.27783203124998e-05\\
-33.8738738738739	2.34863281250002e-05\\
-33.7137137137137	2.43408203124999e-05\\
-33.5535535535536	2.49511718749999e-05\\
-33.2332332332332	2.67333984374999e-05\\
-33.0730730730731	2.7685546875e-05\\
-32.9129129129129	2.88085937499998e-05\\
-32.7527527527528	2.95654296874999e-05\\
-32.5925925925926	3.07128906250001e-05\\
-31.7917917917918	3.5986328125e-05\\
-31.6316316316316	3.75244140624997e-05\\
-30.990990990991	4.24560546875e-05\\
-30.03003003003	5.06103515625002e-05\\
-29.7097097097097	5.46386718749996e-05\\
-29.5495495495495	5.64208984374998e-05\\
-29.3893893893894	5.85693359374999e-05\\
-29.2292292292292	6.00585937500002e-05\\
-28.9089089089089	6.45507812499996e-05\\
-28.7487487487487	6.62597656249995e-05\\
-28.5885885885886	6.84082031250003e-05\\
-28.4284284284284	7.10449218749994e-05\\
-26.6666666666667	9.8291015625e-05\\
-25.7057057057057	0.00011611328125\\
-25.3853853853854	0.0001232666015625\\
-23.7837837837838	0.0001630615234375\\
-23.4634634634635	0.0001721923828125\\
-22.6626626626627	0.0001994873046875\\
-22.3423423423423	0.000210107421874999\\
-21.5415415415415	0.000241015625\\
-20.9009009009009	0.000268139648437502\\
-20.4204204204204	0.0002906005859375\\
-20.1001001001001	0.000305444335937502\\
-19.7797797797798	0.000321728515625001\\
-19.2992992992993	0.0003475830078125\\
-18.978978978979	0.000366992187499998\\
-18.4984984984985	0.000395971679687501\\
-17.8578578578579	0.000439038085937497\\
-17.2172172172172	0.000485546874999998\\
-16.8968968968969	0.000511254882812499\\
-14.1741741741742	0.000780834960937501\\
-8.40840840840841	0.00179763183593749\\
-6.8068068068068	0.0022358154296875\\
-6.006006006006	0.00248422851562498\\
-4.88488488488488	0.00288530273437499\\
-1.52152152152152	0.00440319824218751\\
-0.72072072072072	0.00484118652343751\\
1.84184184184184	0.00644396972656254\\
3.6036036036036	0.00768591308593744\\
5.2052052052052	0.00892468261718755\\
6.8068068068068	0.0102177978515625\\
8.56856856856857	0.0116165039062499\\
9.52952952952953	0.0123632080078125\\
11.1311311311311	0.0135053222656251\\
11.7717717717718	0.0139322753906251\\
13.053053053053	0.0146914306640626\\
14.1741741741742	0.0152627197265624\\
16.8968968968969	0.0161261718750001\\
17.8578578578579	0.01627490234375\\
18.8188188188188	0.0163343749999999\\
20.1001001001001	0.016279443359375\\
21.2212212212212	0.01609951171875\\
22.5025025025025	0.0157938720703124\\
23.7837837837838	0.0153579589843749\\
25.2252252252252	0.01468779296875\\
26.6666666666667	0.0139239501953124\\
27.7877877877878	0.013269677734375\\
29.3893893893894	0.01223017578125\\
30.990990990991	0.0111288085937499\\
32.2722722722723	0.010211328125\\
33.3933933933934	0.00941538085937494\\
34.3543543543544	0.00874418945312495\\
36.4364364364364	0.00735629882812494\\
38.038038038038	0.00635056152343747\\
38.998998998999	0.00578835449218753\\
41.5615615615616	0.00445371093749998\\
43.1631631631632	0.00373818359375003\\
43.8038038038038	0.00348015136718747\\
44.7647647647648	0.00311374511718752\\
45.8858858858859	0.00272719726562499\\
48.2882882882883	0.00201833496093749\\
50.6906906906907	0.00147736816406249\\
51.4914914914915	0.00132595214843749\\
52.2922922922923	0.00118774414062499\\
53.0930930930931	0.001057421875\\
53.8938938938939	0.000942895507812495\\
55.6556556556557	0.000720410156249996\\
56.1361361361361	0.000670922851562502\\
59.019019019019	0.000424340820312498\\
59.8198198198198	0.000371582031249998\\
60.1401401401401	0.000350805664062499\\
};
\addlegendentry{$N=12/20/12, M=20$}

\addplot [color=mycolor2, line width=1.5pt]
  table[row sep=crcr]{%
-40.1201201201201	5.59082031250001e-06\\
-39.95995995996	5.76171874999995e-06\\
-39.7997997997998	5.98144531249999e-06\\
-39.6396396396396	6.10351562499999e-06\\
-39.4794794794795	6.32324218749999e-06\\
-39.3193193193193	6.591796875e-06\\
-39.1591591591592	6.83593749999999e-06\\
-38.998998998999	7.177734375e-06\\
-38.8388388388388	7.44628906250005e-06\\
-38.1981981981982	8.39843750000004e-06\\
-38.038038038038	8.69140624999996e-06\\
-37.8778778778779	9.25292968750002e-06\\
-37.3973973973974	1.0302734375e-05\\
-37.2372372372372	1.04736328125e-05\\
-37.0770770770771	1.07666015625e-05\\
-36.7567567567568	1.15478515625e-05\\
-36.5965965965966	1.18408203124999e-05\\
-36.2762762762763	1.28173828125e-05\\
-36.1161161161161	1.30859374999999e-05\\
-35.955955955956	1.35986328125001e-05\\
-35.7957957957958	1.42578125e-05\\
-35.6356356356356	1.46484375e-05\\
-35.4754754754755	1.53076171875e-05\\
-35.3153153153153	1.61132812499999e-05\\
-35.1551551551552	1.65527343750001e-05\\
-34.994994994995	1.71874999999999e-05\\
-34.8348348348348	1.7724609375e-05\\
-34.5145145145145	1.91162109374999e-05\\
-34.3543543543544	1.95556640624999e-05\\
-34.034034034034	2.09228515624998e-05\\
-33.8738738738739	2.15820312499999e-05\\
-33.7137137137137	2.24121093750001e-05\\
-33.5535535535536	2.34130859375001e-05\\
-33.3933933933934	2.4072265625e-05\\
-33.2332332332332	2.52441406249998e-05\\
-33.0730730730731	2.60009765625001e-05\\
-32.9129129129129	2.65380859375001e-05\\
-32.7527527527528	2.75146484375e-05\\
-32.5925925925926	2.83691406250001e-05\\
-32.4324324324324	2.95166015625001e-05\\
-31.951951951952	3.23486328124999e-05\\
-31.4714714714715	3.57666015624998e-05\\
-31.3113113113113	3.65478515625e-05\\
-31.1511511511512	3.76464843749998e-05\\
-30.990990990991	3.90136718750002e-05\\
-30.1901901901902	4.56298828124998e-05\\
-30.03003003003	4.73144531249998e-05\\
-29.7097097097097	5e-05\\
-29.3893893893894	5.31005859374996e-05\\
-28.9089089089089	5.82275390624998e-05\\
-28.4284284284284	6.39892578124999e-05\\
-27.1471471471471	7.92236328125e-05\\
-26.8268268268268	8.45214843749998e-05\\
-26.1861861861862	9.45312500000001e-05\\
-25.5455455455455	0.0001068359375\\
-24.4244244244244	0.000129809570312501\\
-23.3033033033033	0.000157788085937499\\
-22.3423423423423	0.0001879638671875\\
-21.0610610610611	0.000234179687500001\\
-20.5805805805806	0.000253857421875\\
-19.6196196196196	0.000300122070312498\\
-19.1391391391391	0.000323779296874999\\
-18.6586586586587	0.0003534423828125\\
-18.3383383383383	0.000372314453125002\\
-15.2952952952953	0.000602319335937499\\
-14.1741741741742	0.000715551757812504\\
-12.7327327327327	0.000895434570312495\\
-10.8108108108108	0.00118947753906249\\
-9.20920920920921	0.00149311523437499\\
-8.56856856856857	0.00163581542968749\\
-7.28728728728728	0.001951806640625\\
-5.68568568568568	0.00241894531249999\\
-4.24424424424424	0.00291137695312499\\
-0.880880880880881	0.00441169433593749\\
1.2012012012012	0.00557099609374999\\
4.24424424424424	0.0075732666015625\\
6.48648648648648	0.00925561523437498\\
9.04904904904905	0.0112826416015626\\
10.3303303303303	0.0122543212890626\\
11.6116116116116	0.0131770751953125\\
12.8928928928929	0.014025537109375\\
14.014014014014	0.0146961181640626\\
16.2562562562563	0.0157892578125001\\
17.2172172172172	0.0161342041015625\\
18.1781781781782	0.0164093505859376\\
19.4594594594595	0.0166217773437501\\
20.4204204204204	0.0166600097656251\\
21.3813813813814	0.0166052246093749\\
22.6626626626627	0.01637978515625\\
24.4244244244244	0.0158234863281251\\
26.1861861861862	0.0150476074218751\\
27.6276276276276	0.0142825927734375\\
29.0690690690691	0.0133518310546874\\
30.6706706706707	0.0122286376953125\\
33.0730730730731	0.0104498779296874\\
34.1941941941942	0.00962958984375005\\
35.7957957957958	0.00846669921875005\\
37.0770770770771	0.00757482910156251\\
39.1591591591592	0.00624543457031251\\
41.0810810810811	0.00514555664062503\\
42.2022022022022	0.00457299804687501\\
43.6436436436436	0.00389978027343751\\
44.9249249249249	0.00336391601562497\\
46.3663663663664	0.00283068847656248\\
47.4874874874875	0.00246274414062501\\
48.1281281281281	0.00227163085937501\\
49.4094094094094	0.0019247802734375\\
51.1711711711712	0.0015148193359375\\
51.971971971972	0.00134611816406249\\
53.0930930930931	0.00114584960937501\\
53.5735735735736	0.00106499023437499\\
54.5345345345345	0.00091884765625\\
56.1361361361361	0.000720288085937495\\
57.0970970970971	0.000622949218749996\\
60.1401401401401	0.000382470703124998\\
};
\addlegendentry{$N=12/20/12, M=1$}

\addplot [color=mycolor3, line width=1.5pt]
  table[row sep=crcr]{%
-40.1201201201201	5.10253906249999e-06\\
-39.95995995996	5.37109374999995e-06\\
-39.6396396396396	5.83496093749998e-06\\
-39.4794794794795	5.95703125000004e-06\\
-39.3193193193193	6.20117187500002e-06\\
-39.1591591591592	6.49414062500004e-06\\
-38.998998998999	6.73828125e-06\\
-38.8388388388388	6.86035156249996e-06\\
-38.6786786786787	6.93359375000004e-06\\
-38.038038038038	7.86132812499999e-06\\
-37.8778778778779	8.15429687500002e-06\\
-37.7177177177177	8.56933593749993e-06\\
-37.2372372372372	9.375e-06\\
-37.0770770770771	9.64355468749991e-06\\
-36.7567567567568	1.00341796875e-05\\
-36.4364364364364	1.07177734375e-05\\
-36.2762762762763	1.09619140625001e-05\\
-36.1161161161161	1.1376953125e-05\\
-35.955955955956	1.16943359374999e-05\\
-35.7957957957958	1.21826171875e-05\\
-35.4754754754755	1.30615234375e-05\\
-35.3153153153153	1.33544921875e-05\\
-34.5145145145145	1.56005859375001e-05\\
-34.1941941941942	1.63330078125001e-05\\
-34.034034034034	1.689453125e-05\\
-33.8738738738739	1.73828124999999e-05\\
-33.7137137137137	1.80175781250001e-05\\
-33.5535535535536	1.83837890624999e-05\\
-33.3933933933934	1.9140625e-05\\
-33.0730730730731	2.0263671875e-05\\
-32.9129129129129	2.08251953124999e-05\\
-32.7527527527528	2.1630859375e-05\\
-32.1121121121121	2.45117187499999e-05\\
-31.951951951952	2.52685546875e-05\\
-31.7917917917918	2.61718749999998e-05\\
-31.3113113113113	2.85400390625001e-05\\
-31.1511511511512	2.95654296874999e-05\\
-30.990990990991	3.03955078125001e-05\\
-30.6706706706707	3.25439453125e-05\\
-30.3503503503504	3.43994140624997e-05\\
-30.03003003003	3.65966796874997e-05\\
-29.7097097097097	3.86718749999997e-05\\
-29.5495495495495	4.00878906249999e-05\\
-29.3893893893894	4.1357421875e-05\\
-29.2292292292292	4.23095703125e-05\\
-29.0690690690691	4.38964843750003e-05\\
-28.9089089089089	4.521484375e-05\\
-28.7487487487487	4.63623046874998e-05\\
-28.4284284284284	4.9365234375e-05\\
-28.2682682682683	5.12207031249998e-05\\
-27.7877877877878	5.58349609375003e-05\\
-27.6276276276276	5.73974609374999e-05\\
-27.3073073073073	6.13769531250002e-05\\
-26.6666666666667	6.94335937499995e-05\\
-26.3463463463463	7.31201171874994e-05\\
-25.7057057057057	8.23730468749994e-05\\
-25.3853853853854	8.66210937499999e-05\\
-25.2252252252252	8.90136718750006e-05\\
-25.0650650650651	9.09179687500006e-05\\
-24.5845845845846	9.82421875000007e-05\\
-24.4244244244244	0.000100317382812501\\
-23.9439439439439	0.000109375\\
-23.4634634634635	0.000118334960937499\\
-23.1431431431431	0.0001254150390625\\
-22.6626626626627	0.00013525390625\\
-22.5025025025025	0.000139428710937501\\
-21.8618618618619	0.000153955078124999\\
-21.2212212212212	0.000171289062499999\\
-20.7407407407407	0.0001859130859375\\
-20.1001001001001	0.000206689453125001\\
-19.7797797797798	0.000218725585937501\\
-18.6586586586587	0.000263330078124999\\
-14.1741741741742	0.000539721679687498\\
-13.8538538538538	0.000567968749999998\\
-11.1311311311311	0.000859863281250005\\
-7.44744744744744	0.00148173828125\\
-4.88488488488488	0.00211513671875\\
-4.56456456456456	0.00220437011718749\\
-3.44344344344344	0.00257207031250001\\
-1.2012012012012	0.00341601562500002\\
-0.24024024024024	0.0038416748046875\\
0.880880880880881	0.00439829101562498\\
2.48248248248248	0.00528459472656251\\
3.6036036036036	0.00595117187500004\\
4.72472472472472	0.00666865234375004\\
6.006006006006	0.00751713867187501\\
8.24824824824825	0.00907268066406243\\
9.52952952952953	0.00998793945312496\\
11.7717717717718	0.0115690429687499\\
13.053053053053	0.0123916259765624\\
14.8148148148148	0.013379736328125\\
16.0960960960961	0.01400966796875\\
17.5375375375375	0.0145463623046875\\
18.3383383383383	0.0147924560546874\\
19.4594594594595	0.0150026611328126\\
20.5805805805806	0.0151090332031249\\
22.5025025025025	0.0150552490234376\\
23.6236236236236	0.0148671630859376\\
24.5845845845846	0.0146084228515626\\
26.026026026026	0.0140566162109376\\
27.4674674674675	0.0133866943359374\\
28.7487487487488	0.0127117675781249\\
30.03003003003	0.0119445068359375\\
30.990990990991	0.011325390625\\
32.1121121121121	0.0105844970703125\\
33.7137137137137	0.00951281738281245\\
35.3153153153153	0.00844362792968756\\
36.7567567567568	0.00750915527343748\\
37.8778778778779	0.00682399902343751\\
38.998998998999	0.00616503906249997\\
40.9209209209209	0.00510349121093746\\
42.5225225225225	0.00431708984374998\\
44.2842842842843	0.003541650390625\\
46.2062062062062	0.00281777343750001\\
48.1281281281281	0.00220739746093751\\
48.9289289289289	0.00198566894531251\\
49.2492492492493	0.00190407714843751\\
50.2102102102102	0.00166813964843751\\
50.6906906906907	0.001561669921875\\
51.8118118118118	0.001335400390625\\
52.4524524524525	0.0012197021484375\\
55.015015015015	0.000838110351562499\\
58.6986986986987	0.000466845703124996\\
59.3393393393393	0.000418969726562501\\
60.1401401401401	0.000365576171874997\\
};
\addlegendentry{$N=8/8/8, M=20$}

\addplot [color=mycolor4, line width=1.5pt]
  table[row sep=crcr]{%
-40.1201201201201	3.83300781249997e-06\\
-39.95995995996	3.9306640625e-06\\
-39.7997997997998	4.1015625e-06\\
-39.6396396396396	4.1259765625e-06\\
-39.4794794794795	4.29687499999996e-06\\
-39.3193193193193	4.51660156249999e-06\\
-38.998998998999	4.85839843749999e-06\\
-38.8388388388388	4.93164062500002e-06\\
-38.3583583583584	5.37109374999995e-06\\
-38.1981981981982	5.63964843749995e-06\\
-37.5575575575576	6.44531249999995e-06\\
-37.3973973973974	6.591796875e-06\\
-37.2372372372372	6.78710937499996e-06\\
-37.0770770770771	7.03125e-06\\
-36.9169169169169	7.15332031249997e-06\\
-36.7567567567568	7.51953124999994e-06\\
-36.4364364364364	7.98339843749996e-06\\
-36.1161161161161	8.27636718750003e-06\\
-35.955955955956	8.69140624999996e-06\\
-35.7957957957958	8.93554687499992e-06\\
-35.6356356356356	9.39941406249992e-06\\
-35.4754754754755	9.59472656249993e-06\\
-34.8348348348348	1.10107421874999e-05\\
-34.6746746746747	1.15966796875001e-05\\
-34.3543543543544	1.25732421875e-05\\
-34.1941941941942	1.28662109375e-05\\
-34.034034034034	1.328125e-05\\
-33.7137137137137	1.40625e-05\\
-33.5535535535536	1.45263671875e-05\\
-33.3933933933934	1.52099609374999e-05\\
-33.2332332332332	1.57714843749999e-05\\
-33.0730730730731	1.6455078125e-05\\
-32.9129129129129	1.73095703124998e-05\\
-32.7527527527528	1.78955078125001e-05\\
-32.5925925925926	1.82373046875001e-05\\
-32.4324324324324	1.88720703124999e-05\\
-32.2722722722723	1.92626953124999e-05\\
-31.951951951952	2.05322265625001e-05\\
-31.7917917917918	2.10205078125001e-05\\
-31.6316316316316	2.18261718749999e-05\\
-31.4714714714715	2.2509765625e-05\\
-31.3113113113113	2.35595703124998e-05\\
-30.990990990991	2.49511718749999e-05\\
-30.8308308308308	2.58789062499999e-05\\
-30.6706706706707	2.65869140625e-05\\
-30.5105105105105	2.74414062499998e-05\\
-30.1901901901902	2.94921875000002e-05\\
-30.03003003003	3.04199218749999e-05\\
-29.8698698698699	3.1640625e-05\\
-29.5495495495495	3.37890624999999e-05\\
-29.3893893893894	3.45703124999997e-05\\
-28.1081081081081	4.38232421874996e-05\\
-27.6276276276276	4.82666015624996e-05\\
-25.8658658658659	6.51855468749997e-05\\
-25.7057057057057	6.67480468749998e-05\\
-25.5455455455455	6.86523437500004e-05\\
-25.3853853853854	7.11669921875004e-05\\
-24.4244244244244	8.40576171875006e-05\\
-24.2642642642643	8.61572265624995e-05\\
-24.1041041041041	8.91845703125002e-05\\
-22.8228228228228	0.000112109375\\
-20.1001001001001	0.0001821533203125\\
-19.6196196196196	0.000197119140625\\
-18.4984984984985	0.000236499023437498\\
-18.018018018018	0.000257202148437501\\
-16.5765765765766	0.000325122070312498\\
-16.4164164164164	0.000334423828124998\\
-16.0960960960961	0.000351293945312497\\
-15.4554554554554	0.000388110351562498\\
-15.1351351351351	0.000408862304687497\\
-14.8148148148148	0.000430615234375003\\
-14.1741741741742	0.000476196289062501\\
-13.8538538538538	0.000500976562499996\\
-13.3733733733734	0.000540185546875004\\
-12.8928928928929	0.000581713867187501\\
-10.3303303303303	0.000863745117187501\\
-8.56856856856857	0.00111315917968749\\
-5.04504504504504	0.0018052978515625\\
-4.08408408408408	0.002047021484375\\
-2.48248248248248	0.00251481933593751\\
-2.16216216216216	0.0026253173828125\\
1.52152152152152	0.00411162109374998\\
2.96296296296296	0.00484672851562497\\
5.84584584584584	0.00653166503906245\\
7.44744744744744	0.00759025878906251\\
10.3303303303303	0.00972209472656247\\
12.7327327327327	0.0115025390625001\\
14.8148148148148	0.0129359375\\
16.4164164164164	0.0138833007812501\\
17.6976976976977	0.0144948974609375\\
18.6586586586587	0.0148485839843749\\
20.2602602602603	0.0152862792968751\\
21.7017017017017	0.0154767578125\\
22.6626626626627	0.0154832275390626\\
24.5845845845846	0.01517216796875\\
26.1861861861862	0.0146646484374999\\
27.3073073073073	0.0141947753906249\\
28.7487487487488	0.0134431884765625\\
30.6706706706707	0.0123247070312499\\
31.6316316316316	0.0117032714843749\\
32.9129129129129	0.010809375\\
34.3543543543544	0.00979606933593753\\
35.6356356356356	0.00890275878906242\\
36.5965965965966	0.00825266113281244\\
37.8778778778779	0.00741579589843749\\
38.5185185185185	0.007013623046875\\
40.1201201201201	0.00606464843749997\\
41.5615615615616	0.00530085449218751\\
43.6436436436436	0.004295849609375\\
44.6046046046046	0.00385925292968749\\
45.5655655655656	0.00346057128906249\\
46.3663663663664	0.00314782714843749\\
48.2882882882883	0.00248884277343749\\
48.9289289289289	0.002293017578125\\
52.6126126126126	0.0014096923828125\\
54.5345345345345	0.001069482421875\\
55.4954954954955	0.000927050781249998\\
55.975975975976	0.000863183593749997\\
58.3783783783784	0.000592846679687499\\
59.019019019019	0.000536083984375001\\
59.4994994994995	0.000497119140624997\\
60.1401401401401	0.000446997070312498\\
};
\addlegendentry{$N=8/8/8, M=1$}

\end{axis}
\end{tikzpicture}%

%% file: img/eigenValues.tex
%
%
\definecolor{mycolor1}{rgb}{0.00000,0.44700,0.74100}%
\definecolor{mycolor4}{rgb}{0.49400,0.18400,0.55600}%
\begin{tikzpicture}
\pgfplotsset{every tick label/.append style={font=\scriptsize}}

\begin{axis}[%
width=0.951\fwidth,
height=\fheight,
at={(0\fwidth,0\fheight)},
scale only axis,
xmin=1,
xmax=5,
xtick={1, 2, 3, 4, 5},
xlabel style={font=\footnotesize\color{white!15!black}},
xlabel={Singular Value Index},
ymin=0,
ymax=0.6,
ylabel style={font=\footnotesize\color{white!15!black}},
ylabel={Singular Value Ratio},
axis background/.style={fill=white},
xmajorgrids,
ymajorgrids,
legend style={legend cell align=left, align=left, font=\footnotesize, draw=white!15!black}
]
\addplot [color=mycolor1, line width=1.5pt, mark=*, mark options={solid, fill=mycolor1, mycolor1}]
  table[row sep=crcr]{%
1	0.501796842219875\\
2	0.21217203664956\\
3	0.113905003172795\\
4	0.066451623242987\\
5	0.0405036474706395\\
6	0.024989087874322\\
7	0.0155649673963207\\
8	0.00971751254356572\\
9	0.00596693131876042\\
10	0.00371283580428159\\
11	0.00228097923867916\\
12	0.00135920340454137\\
13	0.000784312695550503\\
14	0.0004494316345158\\
15	0.000239888526977604\\
16	0.000105696806628352\\
};
\addlegendentry{$N=12/20/12$, $M = 20$}

\addplot [color=mycolor4, line width=1.5pt, mark=*, dashed, mark options={solid, fill=mycolor4, mycolor4}]
  table[row sep=crcr]{%
1	0.579298299907249\\
2	0.228108270484178\\
3	0.10589125537536\\
4	0.0498965832465962\\
5	0.022933348728735\\
6	0.0096024842197174\\
7	0.00344448424834166\\
8	0.000821268567499145\\
9	4.00522232288628e-06\\
10	3.96763260840792e-17\\
11	3.09098754782033e-17\\
12	2.72867326877126e-17\\
13	2.46231960879186e-17\\
14	2.2264412087487e-17\\
15	1.99536350525657e-17\\
16	1.7418539486921e-17\\
};
\addlegendentry{$N=8/8/8$, $M = 1$}

\end{axis}
\end{tikzpicture}%